\documentclass{aa}
\usepackage{graphicx}
\usepackage{txfonts}
\usepackage{natbib}
\bibpunct{(}{)}{;}{a}{}{,}

\newcommand{\be}{\begin{enumerate}}
\newcommand{\ee}{\end{enumerate}}

\def\mathstacksym#1#2#3#4#5{\def#1{\mathrel{\hbox to 0pt{\lower#5\hbox{#3}\hss} \raise #4\hbox{#2}}}}
\mathstacksym\gta{$>$}{$\sim$}{1.5pt}{3.5pt} 
\mathstacksym\lta{$<$}{$\sim$}{1.5pt}{3.5pt} 
\newcount\longrefs
\def\aap{\ifnum\longrefs=1 {Astron.\ Astrophys.}\else 
                           {A\hbox{\rm \&}A}\fi}
\def\aaps{\ifnum\longrefs=1 {Astron.\ Astrophys.\ Suppl.}\else 
                            {A\hbox{\rm \&}A Suppl.}\fi}
\def\aj{\ifnum\longrefs=1 {Astron.\ J.}\else 
                          {AJ}\fi} 
\def\apj{\ifnum\longrefs=1 {Astrophys.\ J.}\else 
                           {ApJ}\fi} 
\def\apjs{\ifnum\longrefs=1 {Astrophys.\ J. Suppl.}\else 
                            {ApJS}\fi}
\def\araa{\ifnum\longrefs=1 {Ann.\ Rev.\ Astron.\ Astrophys.}\else 
                            {ARA\hbox{\rm \&}A}\fi}
\def\mnras{\ifnum\longrefs=1 {Mon.\ Not.\ Roy.\ Astron.\ Soc.}\else 
                             {MNRAS}\fi} 
\def\pasp{\ifnum\longrefs=1 {Pub.\ Astron.\ Soc.\ Pacific}\else 
                            {PASP}\fi} 
\begin{document}
%
%
\title{A Quest for PMS candidate stars at low metallicity: Variable HAe/Be and Be stars in the 
Small Magellanic Cloud}
\titlerunning{SMC HAe/Be candidate stars}

\author{        W.J. de Wit\inst{1,2,3}     \and
                J-P. Beaulieu\inst{3}       \and
                H.J.G.L.M. Lamers\inst{1,2} \and
                E. Lesquoy \inst{3}         \and 
                J-B. Marquette \inst{3}  
                } 
\offprints{W.J. de Wit, \email{dewit@arcetri.astro.it}}
\institute{     Astronomical Institute, University of Utrecht,
                Princetonplein 5, NL-3584CC, Utrecht, The Netherlands \and 
                SRON Laboratory for Space Research, Sorbonnelaan 2, NL-3584 CA, 
                Utrecht, The Netherlands                               \and 
                Institut d'Astrophysique de Paris, 98bis Boulevard Arago, 75014 Paris, France}
\date{Received date; accepted date}
\abstract{
We report the discovery of 5 new Herbig Ae/Be  candidate stars in the Small
Magellanic Cloud  in addition to the 2 reported in Beaulieu et
al. (2001). We discuss these 7 HAeBe candidate stars in terms of (1) their
irregular photometric variability, (2) their near infrared emission, (3) their
H$\alpha$ emission and (4) their spectral type. One star has the typical
photometric behaviour that is observed only among Pre-Main Sequence UX\,Orionis
type stars. The objects are more luminous than Galactic HAeBe stars and Large
Magellanic Cloud HAeBe candidates of the same spectral type.
\newline
The stars were discovered in a systematic search for variable stars in a subset of
the EROS2 database consisting of 115\,612 stars in a field of 24x24 arcmin in
the Small Magellanic Cloud. In total we discovered 504
variable stars. After classifying the different objects according to their type of variability,
we concentrate on 7 blue objects with irregular photometric behaviour.
We cross-identified these objects with emission line catalogues from Simbad and
JHK photometry from 2MASS. The analysis is supplemented with obtained narrow and
broad band imaging. We discuss their variability in terms of dust obscuration
and bound-free and free-free emission. We estimate the influence of metallicity
on the circumstellar dust emission from pre-main sequence stars.
}

\maketitle

\section{Introduction}

Can pre-main sequence stars in a low metallicity environment be of a 
higher luminosity?


This question forms the basis of a study involving blue variable stars in the
Large Magellanic Cloud (LMC) by Lamers et
al. (1999)\nocite{1999A&A...341..827L}. These stars were initially identified as
intermediate mass pre-main sequence (PMS) candidates in Beaulieu et
al. (1996)\nocite{1996Sci...272..995B} due to their brightness variability.
This variability resembles most closely the irregular variability observed among
Herbig Ae/Be (HAeBe) stars (Herbig 1960\nocite{1960ApJS....4..337H}, Th\'{e}
1994\nocite{1994nesg.conf.....T}, Waters \& Waelkens
1998\nocite{1998ARA&A..36..233W}, Herbst \& Shevchenko
1999\nocite{1999AJ....118.1043S}). In Lamers et al. (1999) it is shown that
there are more common properties between these stars and HAeBes than just their
variability. To summarize: the stars are (1) B or A type stars with Balmer line
emission (2) of luminosity class III-V (3) associated with \ion{H}{ii} regions (4)
located to the right of the Main Sequence in the HR-diagram.  The authors
subsequently argue for a possible PMS nature, and as such, these objects would
be the first PMS stars identified in the Magellanic Clouds  (but see also
Keller et al. 2002\nocite{2002AJ....124.2039K}). Interestingly, it
was shown that these stars are intrinsically brighter than most of the Galactic
HAeBe stars (see e.g. van Duuren et al. 1994\nocite{1994nesh.conf..193V}, Testi
1998\nocite{1998A&AS..133...81T}). Under the PMS assumption, comparison with the
upper luminosity limit in the HR-diagram (i.e. the birthline, see Palla \&
Stahler 1993) for Galactic PMS stars is readily made. Then these LMC PMS
candidates are ten times more luminous than the generally adopted Galactic
birthline. The sample of the {\bf E}ros {\bf L}MC {\bf H}AeBe {\bf C}andidates
(ELHCs) was extended in De Wit et al. (2002)\nocite{2002A&A...395..829D} by 14
objects. In total the group consists of 21 members of which the large majority
is located above the Galactic birthline in the HR-diagram. This result implies
that in the LMC, we might observe more massive stars in an earlier evolutionary
stage than in the Galaxy.

The birthline in the HR-diagram is determined by the proto-stellar mass
accretion rate.  Therefore the Magellanic Cloud's PMS population will provide
important clues to the physics of proto-stellar mass accretion with respect to
its dependence on metallicity.  An observable metallicity effect will only be
detected by doing an inter-comparison between genuine PMS stars in the Galaxy,
LMC, and SMC.

The discovery of the first two {\bf E}ros {\bf S}MC {\bf H}AeBe {\bf C}andidates
(ESHCs) was presented by Beaulieu et al. (2001)\nocite{ 2001A&A...380..168B}.  The characteristics
of these stars are similar to their counterparts in the LMC, and to features
observed among galactic HAeBe stars. Importantly, the
luminosities of these two stars are comparable to those of the PMS
candidates in the LMC.

In this paper we will present a systematic search for variables in the SMC
using the EROS2 light curve database, with first objective to identify blue
irregular variables. The search resulted in 5 objects in addition to
the 2 objects (ESHC\,1 and 2) discussed by Beaulieu et al. (2001). Many other types of
variables were also found. These will be presented in a forthcoming paper. We
will describe the EROS2 data in Sect.\,\ref{obs}. In the same section additional
photometric observations will be presented. The subject of Sect.\,\ref{hacan} is
the compilation of our sample of SMC PMS candidates. The light and colour behaviour
of these stars will be analysed in Sect.\,\ref{descoflc}. The following two
sections treat the H$\alpha$ emission and the cross identifications with existing archival observations and
catalogues.  Then in Sect.\,\ref{sedfit} we will analyse the Spectral Energy Distribution
of the SMC PMS candidates for which we obtained BVRI broad band
photometry.  In Sect.\,\ref{discussion} we will discuss extensively the resulting
properties of the SMC PMS candidates and compare them with different models and
predictions for stars in a Pre Main-Sequence stage as well as for stars in a Post Main-Sequence stage.
We will evaluate the nature of this sample of irregular variable stars in
Sect.\,\ref{discnat} and present our conclusions in Sect.\,\ref{conclusions}.

\section{Observations}
\label{obs}
\subsection{The EROS2 catalogue}
\label{parobs}
The photometric observations, which form the basis of our search for variable stars 
are the product of the EROS2 microlensing
survey. The set-up consists of a 1m F/5 Ritchey Chr\'etien telescope at ESO La
Silla with two 4k\,x\,8k CCD mosaic in different focal planes.  Observations are
simultaneously done with two broad band filters called $V_{\rm E}$ and $R_{\rm E}$. The
definition of the EROS2 system is that the intrinsic colour of an A0 star has
$V_{\rm E}-R_{\rm E}=0$, while the zero-points are such that a star of zero colour will
have a value for its $V_{\rm E}$ magnitude that will equal its Johnson $V_{\rm J}$
magnitude. The EROS2 SMC data in this paper consists of about 330 images per filter,
 covering 0.45 x 0.45 degrees on the sky.  The covered area is split in
four fields, which have the following EROS2 identifications: sm00101l, sm00101n,
sm00103k, sm00103m.  The coordinates of the fields are given in Table\,1. The
observations stretch a period of 1000 days, from February 1996 to April 1999. 
\begin{table}
 {
 \begin{center}
 \caption[]{Coordinates of the centres of the 4 SMC fields.}
  \begin{tabular}[t]{llll}
  \hline
   Name   &   RA(2000)   &  Dec(2000)   \\
          &   (h,m,s)    &  ($^0,',''$) \\ 
   \hline
   sm00101l  & 00:52:52& -73:10:24\\
   sm00101n  & 00:54:56& -73:10:34\\ 
   sm00103k  & 00:52:50& -73:22:30\\
   sm00103m  & 00:55:00& -73:22:20\\ 
  \hline 
  \end{tabular}
 \end{center}
 }
\label{tabfi}
\end{table}

\subsubsection{EROS2 calibration using OGLE data}
\label{oglecal}
The EROS2 experiment observes in a particular filter system. 
In order to calibrate this system to the standard passbands, 
we cross identified all the EROS2 sources with the publically available
SMC BVI catalogues from OGLE (Udalski et al. 
1998\nocite{1998AcA....48..147U}). 
The available OGLE measurements are an average over the period June 26, 1997
(JD\,2450625.5) through March 4, 1998 (JD\,2450876.5). We used stars with reliable
measurements (ie, $\sigma (V_{\rm OGLE}) < 0.02$ and $\sigma
(V_{\rm E})<0.02$) and excluded stars with deviant
magnitudes. Predominantly these were long period red variable stars,
for which the average magnitude between OGLE and EROS2 do
not agree. We derived second order transformation
equations from the EROS2 system to the standard Johnson-Cousins system for each EROS2 field.
For field sm00101l:
\begin{equation}
(V-I) = 0.00 + 1.26*(V_{E} - R_{\rm E}) + 0.19*(V_{\rm E} - R_{\rm E})^{2}
\end{equation}
\begin{equation}
(V_{\rm E}-V) = 0.00 - 0.17 (V-I) - 0.08 (V-I)^{2}
\end{equation}
For field sm00101n:
\begin{equation}
(V-I)  = 0.00 + 1.25 (V_{\rm E} - R_{\rm E}) + 0.18*(V_{\rm E} - R_{\rm E})^{2}
\end{equation}
\begin{equation}
(V_{\rm E}-V) = 0.03 - 0.17 (V-I) - 0.07 (V-I)^{2}
\end{equation}
For field sm00103k:
\begin{equation}
(V-I) = 0.04 + 1.29 (V_{\rm E} - R_{\rm E}) + 0.21*(V_{\rm E} - R_{\rm E})^{2}
\end{equation}
\begin{equation}
(V_{\rm E}-V) = 0.00 - 0.17 (V-I) - 0.08 (V-I)^{2}
\end{equation}
For field sm00103m:
\begin{equation}
(V-I) = 0.02 + 1.28 (V_{\rm E} - R_{\rm E}) + 0.22*(V_{\rm E} - R_{\rm E})^{2}
\end{equation}
\begin{equation}
(V_{\rm E}-V) = 0.02 - 0.17 (V-I) - 0.08 (V-I)^{2}
\end{equation}

\begin{figure}
\includegraphics[height=8cm,width=8cm]{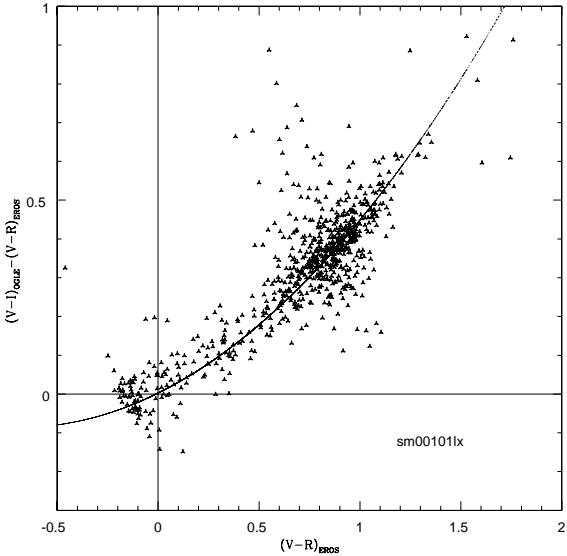}
 \caption[]{Colour calibration for EROS2 passbands of CCD sm00101l, using OGLE data. Plotted are
 the residuals ($(V-I)_{OGLE}-(V-R)_{EROS})$) as function of $(V-R)_{EROS})$.} 
\label{colcal}
\end{figure}

\begin{figure}
\includegraphics[height=8cm,width=8cm]{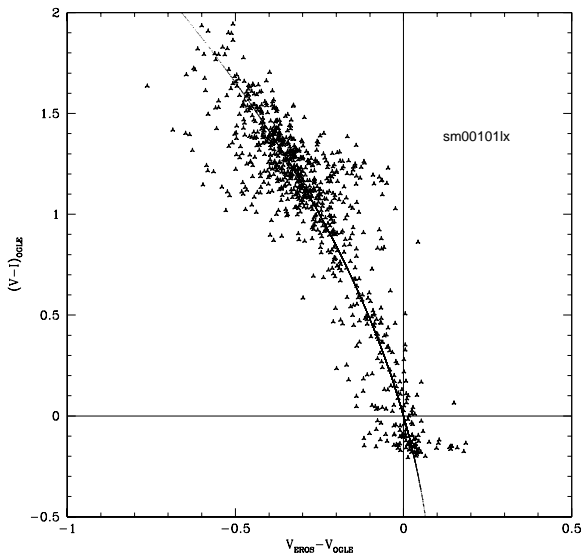}
 \caption{Magnitude calibration for EROS2 passbands of CCD sm00101, using OGLE data. The figure
shows the dependence of the transformation from $V_{\rm E}$ to $V$ on the colour of the star.} 
\label{magcal}
\end{figure}

As an example we give in Figs.\,\ref{colcal} and \ref{magcal} the resulting
relation for one field, viz. sm00101l.  
For this particular field we use 925 stars for the calibration. 

The EROS2 filters are so wide that a colour transformation equation can be
derived only to an uncertainty of $\sigma=0.05$. We stress that
the EROS2 photometry should be used with caution, when deriving stellar
parameters from the observed EROS2 colours alone.

\subsection{ESO BVRIH$\alpha$H$\alpha_{cont}$ Photometry}
\label{danphot}
On January 7, 1998 (JD 2450821.4) we obtained broad and narrow band photometry
at ESO La Silla with DFOSC on the Danish
1.5m\footnote{http://www.ls.eso.org/lasilla/Telescopes/2p2T/D1p5M.} for three
selected SMC fields. One field contained three of the seven ESHCs. We used BVRI
and H$\alpha$H$\alpha_{cont}$ filters with exposure times of 60s, 30s, 30s,
 30s, 600s and 600s, respectively.  The filters are the so-called
ESO filters numbered 450, 451, 452, 425, 693, 697, respectively.  The CCD is a
LORAL 2048x2048 CCD with a pixel scale of 0.39 arcsec/pixel.  The data have been
reduced using QUYLLURWASI, the pipeline developed by one of us (JPB) around
DoPhot for the PLANET (Albrow et al. 1998\nocite{1998ApJ...509..687A})
collaboration.  The absolute calibration has been done using Landolt standards
taken during the observations. This leads to an absolute calibration accuracy of
2\,\% for stars brighter than $V=17^{m}$, and of the order of 5\,\% for stars of
$V=18^{m}$. The field is so crowded that incompleteness starts to be severe at
$V=18^{m}$.  This was determined from the observed distribution of stars on the
main sequence.  Details on the data and data reduction can be found in Stegeman
et al. (2003)\nocite{}.

\section{HAeBe candidates in the EROS2 catalogue}
\label{hacan}
\subsection{Selection and a first classification of variable stars}
\label{selection}
We conducted a systematic search for photometric variability in the EROS2
dataset. We used the one way Analysis of Variance (AoV) method of
\cite{1989MNRAS.241..153S} as the variability search method, following De Wit et
al. (2002).  We searched for variable signals in the period range 0.2 to 1000
days with a frequency step of $\rm 0.00025\,day^{-1}$.  If the peak of the AoV
periodogram is greater than a certain level ($> 25$), a finer frequency mesh
with an oversampling factor of 50 is adopted around the main peak. In total
115\,612 stars were examined for variability in this way.

The AoV method can detect irregular variations or long term modulation more
efficient than other period search methods.  In De Wit et al. (2002) it was
shown that, depending on the characteristics (amplitude, time scale) of the
variation, AoV can produce peaks in the periodogram at the typical time scale of
variation, indicating a significant photometric variation.  This is important,
considering the irregular photometric behaviour of pre-main sequence stars,
semi-regular and irregular type red variables.

Variable stars with a maximum power produced by AoV higher than a certain
threshold level were used for further investigation. The threshold level was
determined empirically from the maximum power distribution as function of the
$R_{\rm E}$ magnitude of all available stars in the EROS2 database. Basically,
fainter stars should produce higher values of their maximum power than brighter
stars in order to be considered a variable object. This takes into account the
larger probability of obtaining a higher maximum power, when the noise level is
higher. A resulting sample of 844 stars was compiled.  The light curves of the
stars in this sample were inspected visually. This enabled us to distinguish
between the different types of variable stars, depending on the period, colour
and shape of their light curve. We initially classify them as
Cepheid-RR\,Lyrae-popII pulsators, Eclipsing Binaries, Long Period Variables
(LPVs), Blue Variables and HAeBe candidates. The total number of each type of
variable in every CCD is listed in Table\,~\ref{totnum}.

\begin{table}
 \centering
 \caption[]{Number of variables in the examined SMC fields.}
  \begin{tabular}[t]{lllll}
   \hline
   EROS2 Fields   &  101l &   101n &  103k & 103m \\
   \hline
   Type of objects:    &    &    &    &   \\
   Pulsators           & 83 & 41 & 67 & 59 \\
   Eclipsing binaries  & 14 & 14 & 16 & 7  \\
   LPVs                & 53 & 41 & 38 & 44 \\
   Blue variables      & 8  &  7 & 5  & 0  \\
   HAeBe candidates    & 2  &  2 & 3  & 0  \\
   \hline
    \end{tabular}
\label{totnum}
\end{table}

\begin{figure}
\centering
\resizebox{\hsize}{!}{\includegraphics[width=11cm,height=12.5cm]{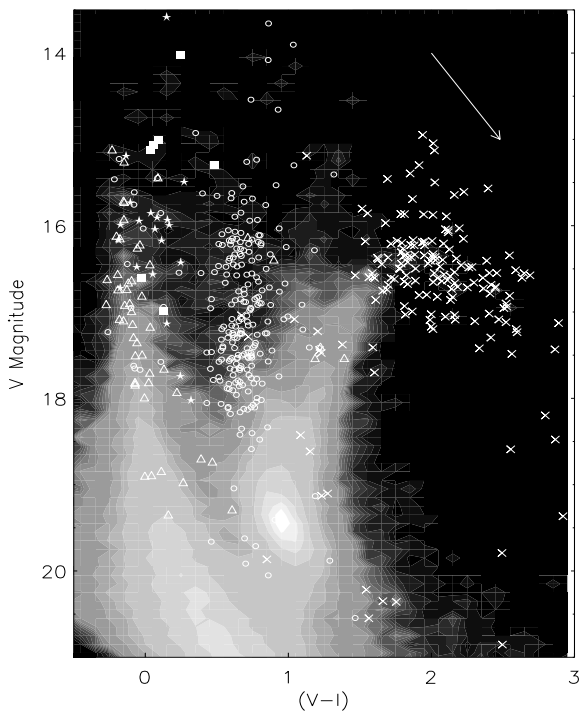}}
 \caption[]{Combined colour magnitude diagram of the 4 target EROS2 SMC fields consisting of 115\,000
 stars. The grey scale represents number of stars. The HAeBe candidates  
are indicated by filled
squares, the eclipsing binaries by triangles, pulsators by circles,
blue variables by asterisks and crosses are both the long period
variables and the red irregular variables.  The arrow indicates the
extinction vector for an extinction of 1 magnitude.}
\label{figCMD}
\end{figure}

In Fig.\,\ref{figCMD} we show a Hess diagram (Hess 1924),\nocite{hess24} which
describes the density distribution of stars in the colour-magnitude plane of the
approximately 115\,000 stars within the examined field. We included the identified
variable stars, which are represented by different symbols. 
All presented variables have been inspected visually. 
Magnitude and colours are in the Johnson-Cousins system,
calibrated with the OGLE SMC data.  The different types of variables have their own
symbol. 

\subsection{Selection of blue irregular variable stars}
The combined number of 27 blue variable stars and HAeBe candidates listed in Table\,\ref{totnum}
are objects with average colour of $(V_{\rm E}-R_{\rm E}) < 0.35$. This colour corresponds
to unreddened Galactic main-sequence (MS) stars of type F2 and earlier. In the
SMC the colour criterion corresponds to MS star of type A8 and earlier, applying
the known foreground extinction towards the SMC of $E(B-V)=0.07$ (Larsen et
al. 2000\nocite{2000A&A...364..455L}).  Among these blue objects a variety of
light curves is present (see also Mennickent et al. 2002\nocite{2002A&A...393..887M}). We will group these stars on the basis of the observed
time scale and amplitude of variations: (A) modulations with distinct periodicity
(2 objects); (B) long term ($\gta100$ days) modulations (7 objects); (C)
intermediate term ($\rm 3\lta days \lta 50$) irregular variations with or
without long term modulations with amplitude $\Delta V_{\rm E}>0.2$ (7
objects); (D) intermediate time scale variation with small amplitude $\Delta
V_{\rm E}<0.1$ (11 objects). 

The expected time scale of brightness changes of the Galactic intermediate mass
PMS stars complies with group (C).  These seven stars we have termed therefore HAeBe
candidates and will be analysed in this paper.  The coordinates, magnitude and
colour of these 7 HAeBe candidates are given in Table\,~\ref{eshctab}.  The
other 20 blue variables will be discussed in a forthcoming paper.
\begin{table*}
\centering
 \caption[]{HAeBe candidates discussed in this paper.}
 \begin{tabular}[t]{cllcccc}
\hline
   Name   &  EROS2-name &  Other design. & RA(2000)   &  Dec(2000) &  $\rm <V>$ & $\rm <V-I> $\\    
   ~~~~~  &             &               &   (h,m,s)  & ($^0,',''$)&        &       \\    
   \hline
\object{ESHC\,1}  &  sm00103k-4755    & LIN 232     & 00 53 02.89  &  -73 17 59.0 & 15.05 & 0.06 \\ 
\object{ESHC\,2}  &  sm00103k-2210    &             & 00 52 32.73  &  -73 17 06.9 & 16.97 & 0.14 \\ 
\object{ESHC\,3}  &  sm00101l-15299   & [MA93] 563  & 00 52 21.59  &  -73 13 32.1 & 16.61 & -0.02\\ 
\object{ESHC\,4}  &  sm00101n-14763   &             & 00 53 56.77  &  -73 10 30.0 & 14.97 & 0.09\\ 
\object{ESHC\,5}  &  sm00101l-3359    & [MA93] 535  & 00 52 06.55  &  -73 06 29.3 & 14.02 & 0.24 \\ 
\object{ESHC\,6}  &  sm00101n-2402    & LIN 264     & 00 54 37.64  &  -73 04 56.4 & 15.09 & 0.04 \\ 
\object{ESHC\,7}  &  sm00103k-6329    & [MA93] 619  & 00 52 52.60  &  -73 18 33.4 & 15.27 & 0.50 \\ 
\hline
   \noalign{\smallskip}
  \end{tabular}
\label{eshctab}
\end{table*}

Two stars of group (C) were studied photometrically and spectroscopically by
Beaulieu et al. (2001). Following their nomenclature we will classify these
stars as {\bf E}ros {\bf S}MC {\bf H}Ae/Be {\bf C}andidates, or ESHCs.

We would like to stress that the terminology used, is based on the variability
similarities with stars discovered in the LMC (the ELHCs) in Lamers et al. (1999) and De Wit et
al. (2002). Some ELHCs have been studied in more detail in LBD.  Considering
that these are their SMC counterparts, we will use the name ESHC, emphasizing that
at present these stars are {\it candidate} PMS stars.

\section{Description of brightness and colour variations of the ESHCs}
\label{descoflc}
\begin{figure*}
\centering
\includegraphics[width=6cm,height=8.5cm,angle=90]{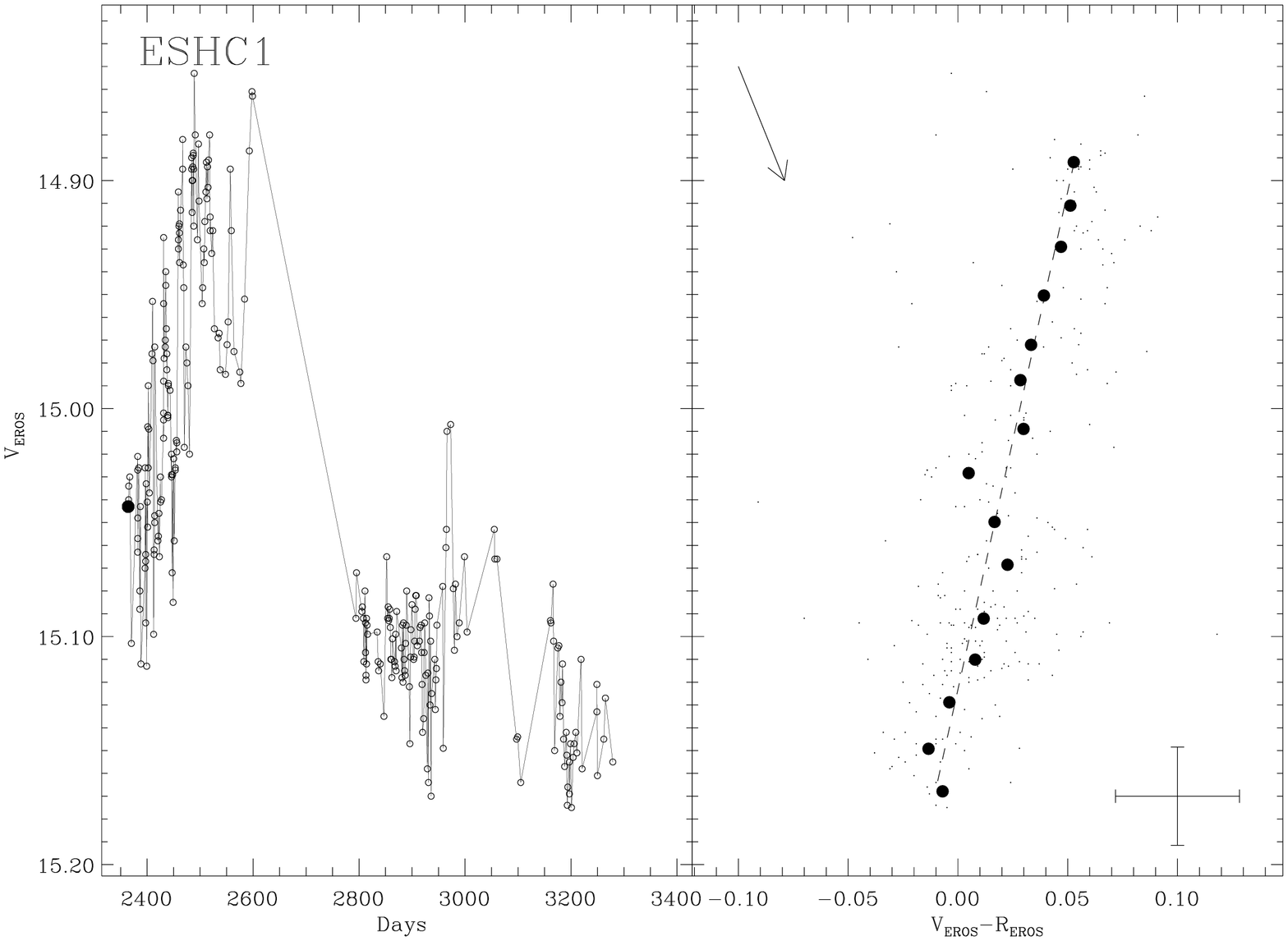}
\includegraphics[width=6cm,height=8.5cm,angle=90]{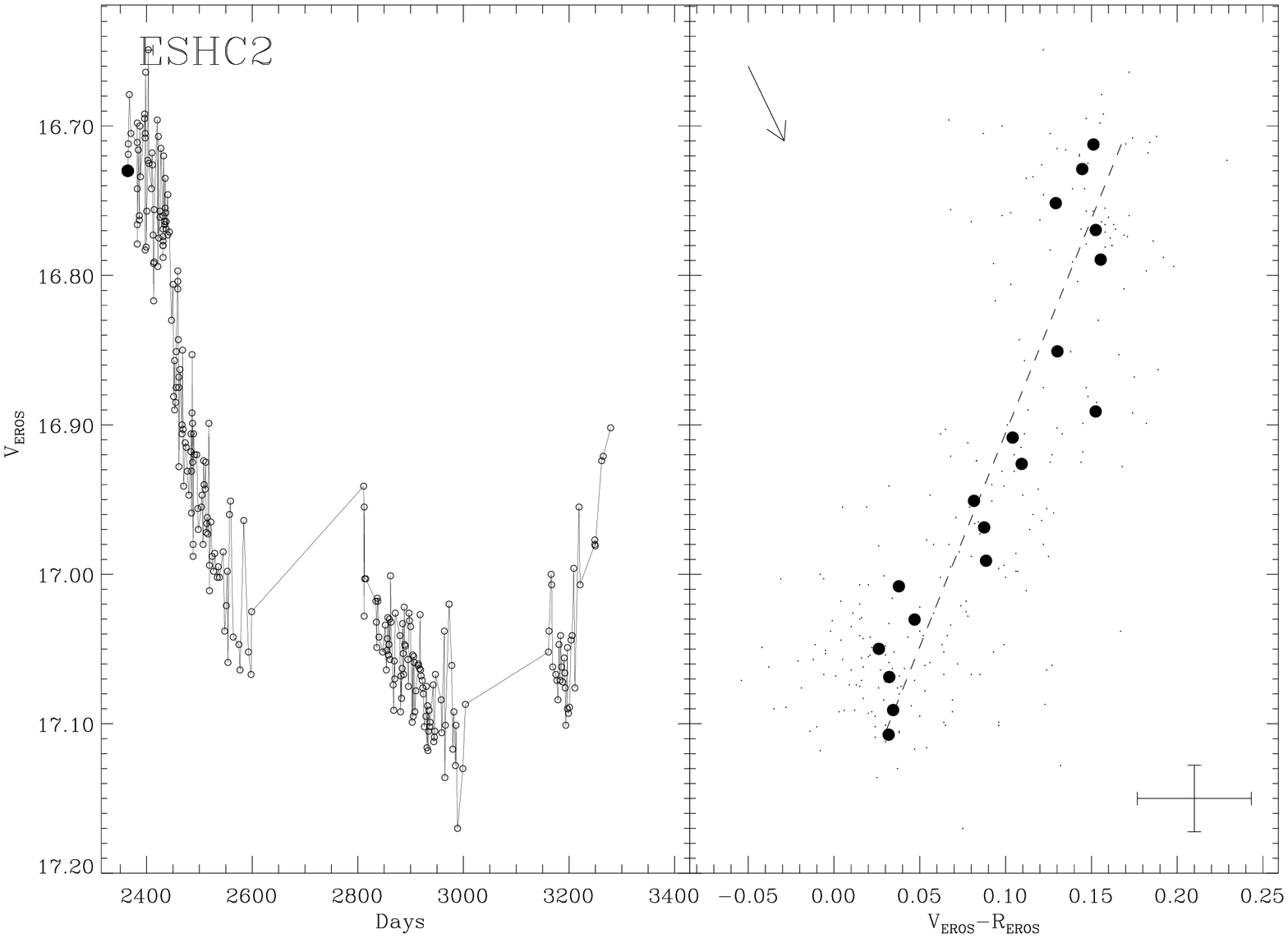}
\includegraphics[width=6cm,height=8.5cm,angle=90]{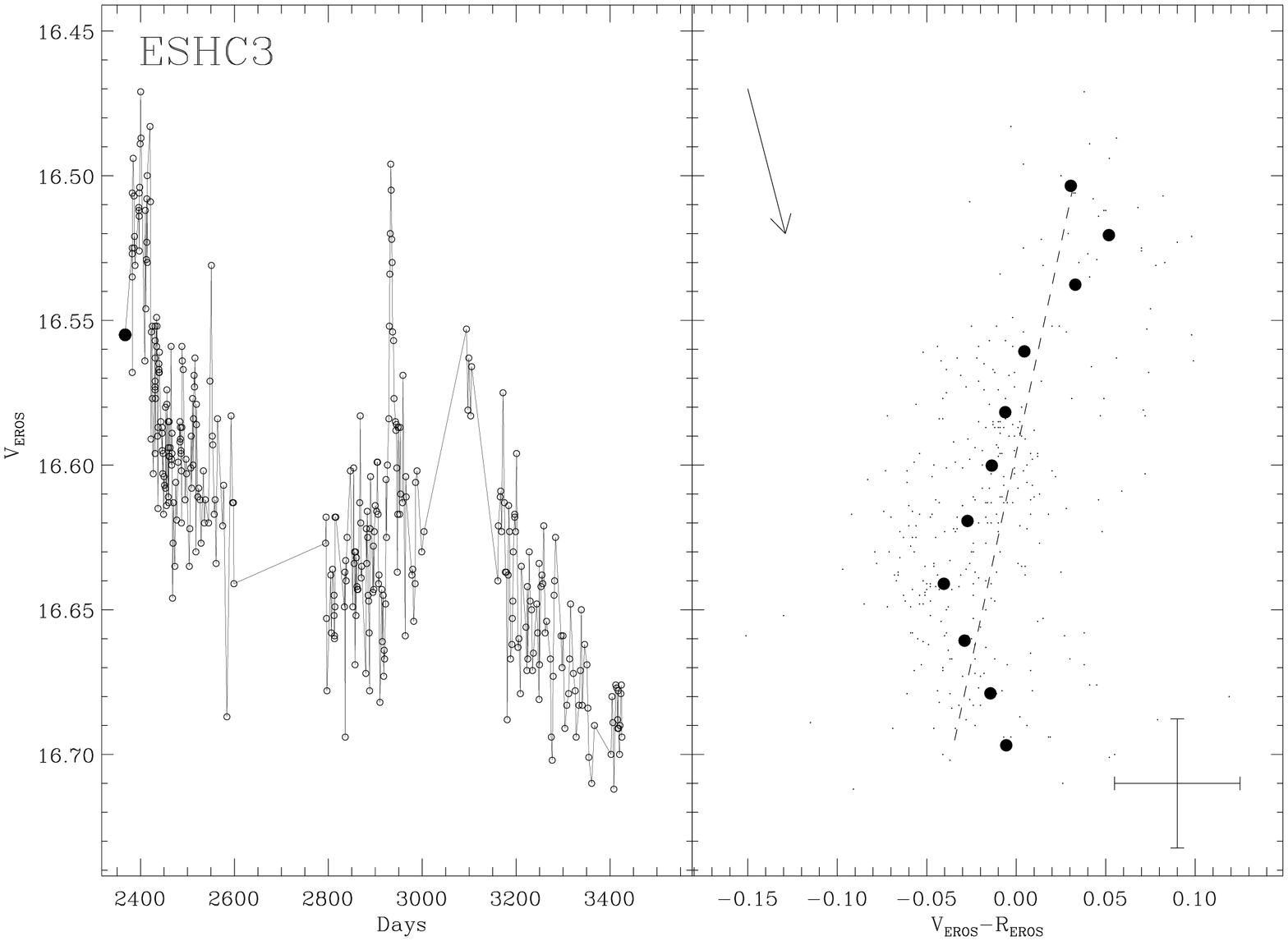}
\includegraphics[width=6cm,height=8.5cm,angle=90]{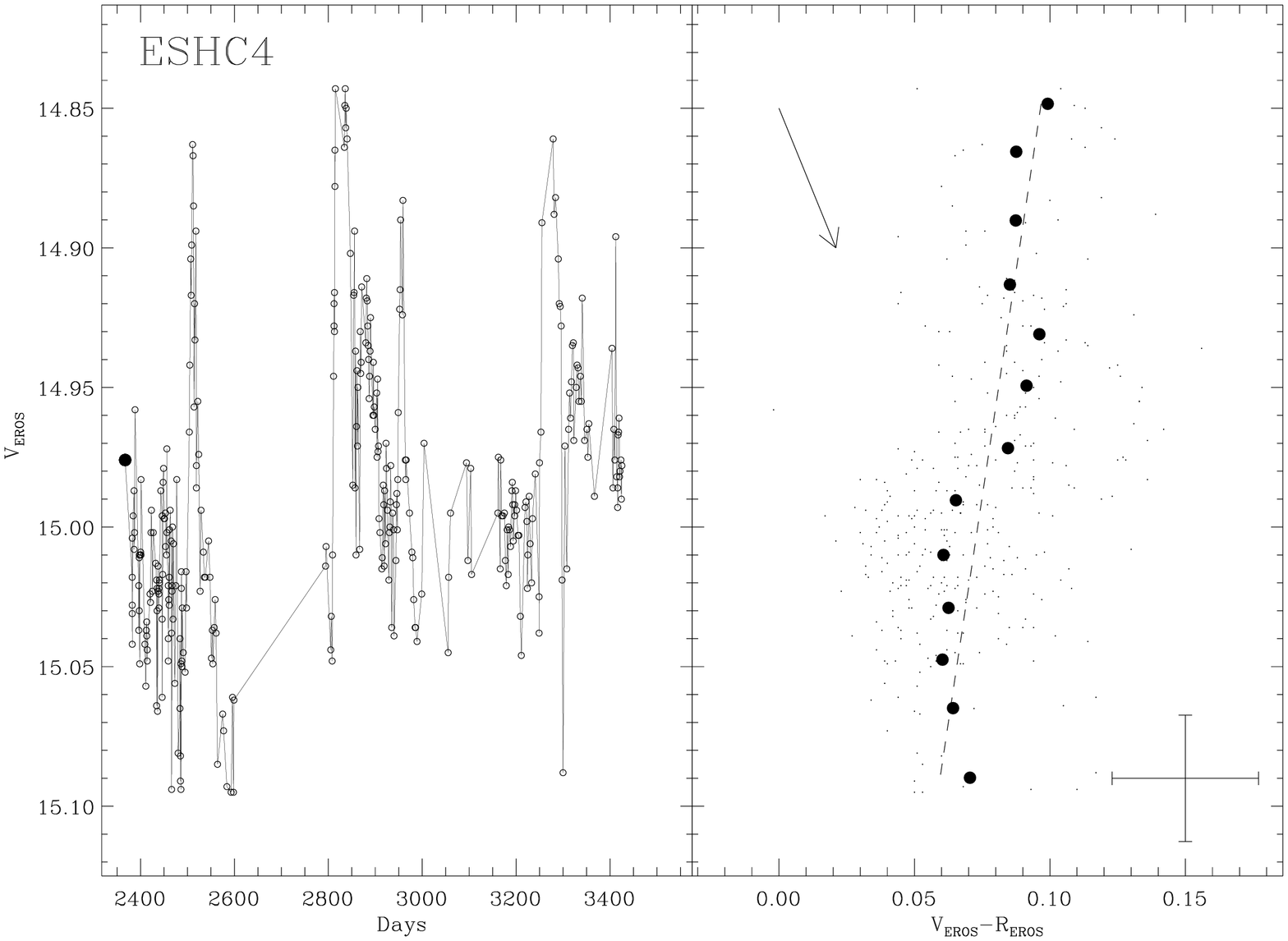}
 \caption{The light curves and corresponding CMDs of ESHC\,1 to 4. The time is given as
 JD-2447892.5 (the time axis runs from 19apr 1996 through 24apr 1999).} 
\label{eshc14lc}
\end{figure*}

\begin{figure*}
\centering
\includegraphics[width=6cm,height=8.5cm,angle=90]{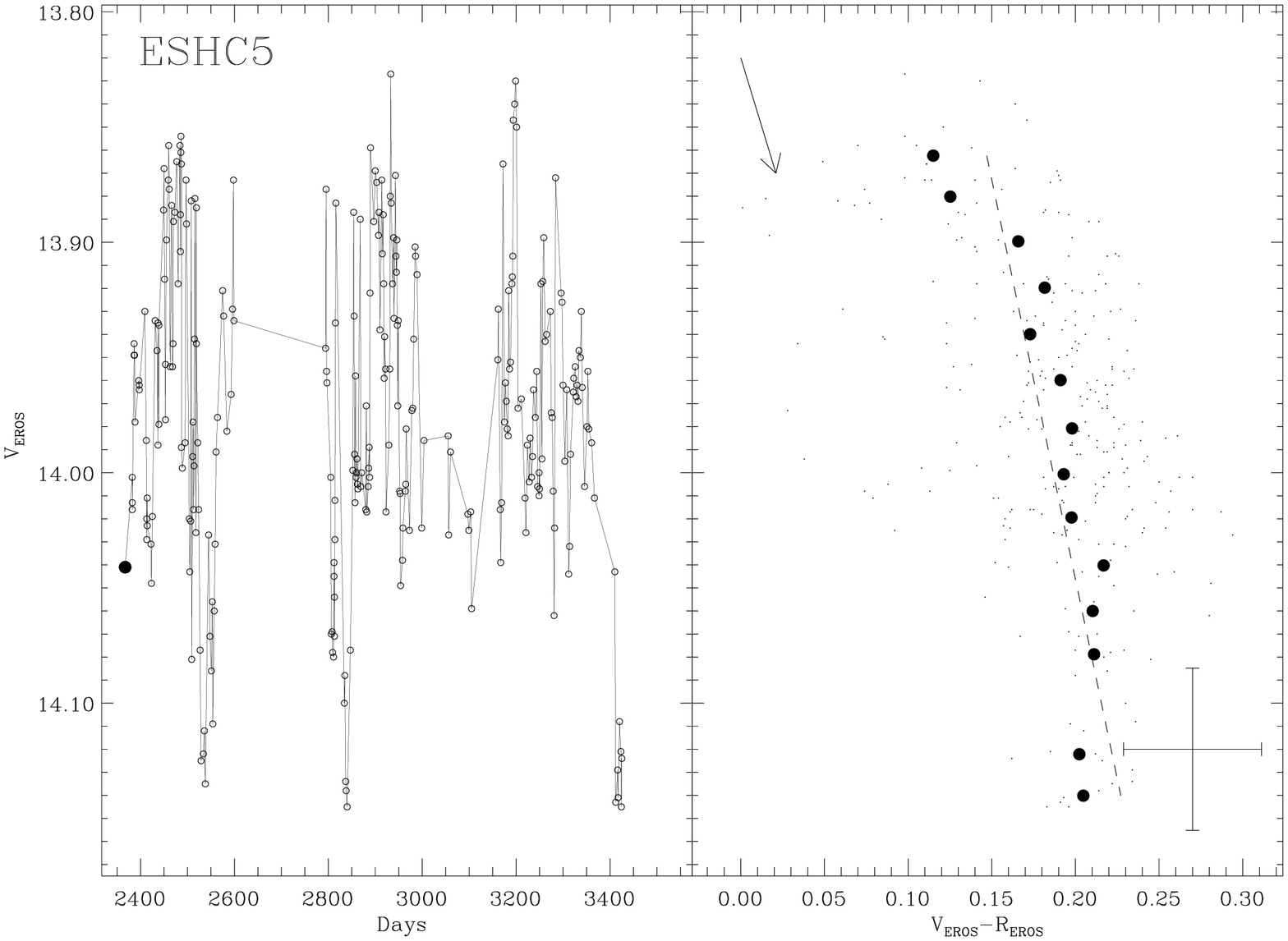}
\includegraphics[width=6cm,height=8.5cm,angle=90]{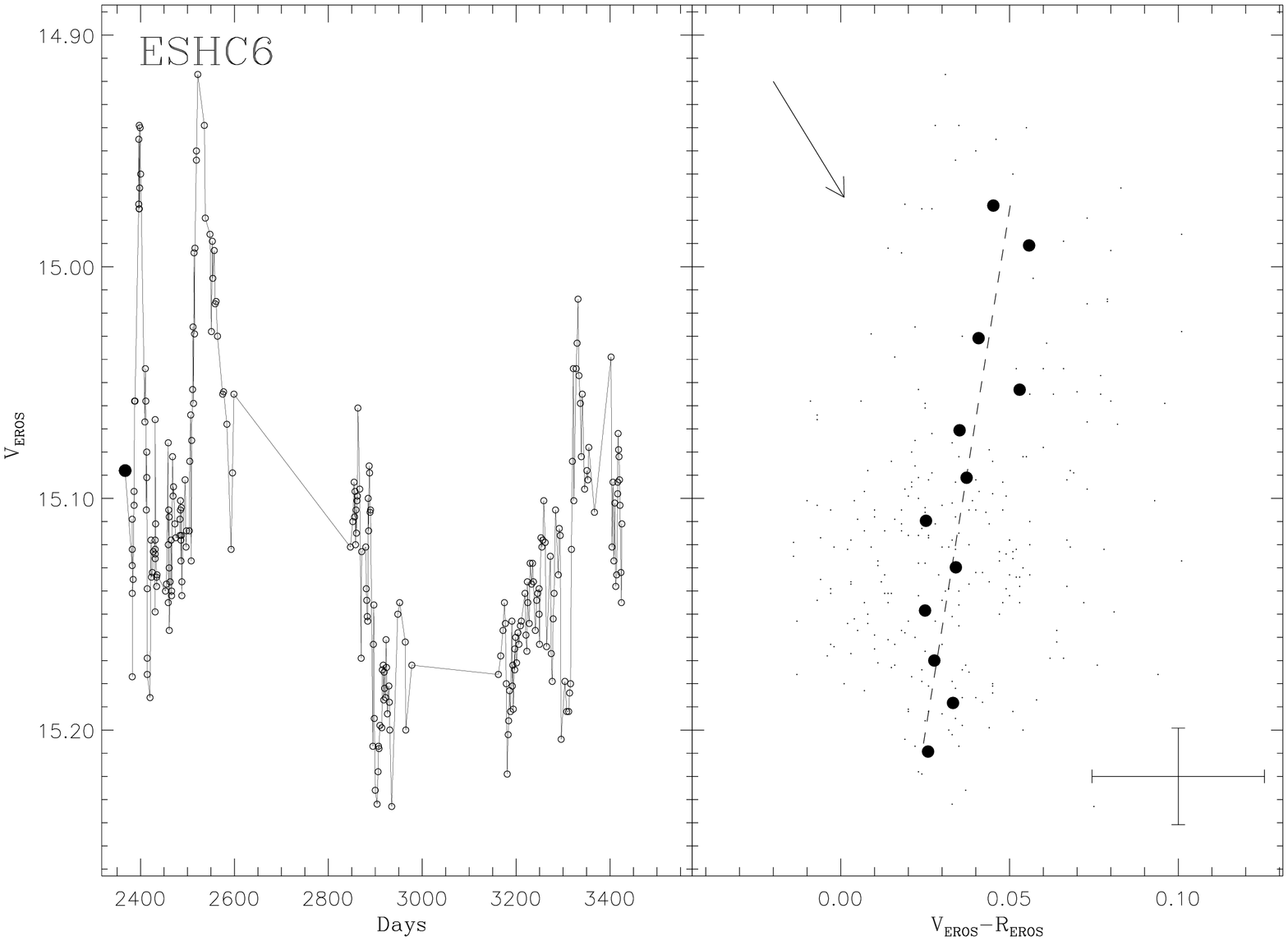}
 \caption{The light curves and corresponding CMDs of ESHC\,5 and 6. The time is given as
 JD-2447892.5 (the time axis runs from 19apr 1996 through 24apr 1999).} 
\label{eshc57lc}
\end{figure*}

The light curves of the ESHCs form the first criterion to identify these objects
as such.  They are presented in Figs.\,\ref{eshc14lc}, \ref{eshc57lc}, and \ref{eshc7cmd}. 
Also presented is the corresponding behaviour of the stars in colour-magnitude
diagram (CMD). The observations span more than 3 years. On average there are
330 measurements per star. The mean time-interval between observations during an
observing season is $\lta 2$ days. The error bars are indicated in the CMDs. 
The EROS2 photometric accuracy for a star of $R_{\rm E}=15^{m}$ is
$\sigma(R_{\rm E})\sim0.02$.  The filled circles in the CMDs are the average
colours in magnitude bins of 0.02m, provided there are 
more than 4 measurements in each magnitude bin. The dashed line is a linear
fit to these points and the value of the slopes is given in Table\,4. These fits
have been put to make clear the observed general trend. In the next
subsections, we will give a short description of the light curves. 


%



\subsection{The variability of ESHC\,1 to 6}
In general the light curves do not show any sign of periodicities in their
signal, but irregular photometric variations on intermediate time scales ($3 \lta
\rm days \lta 50\,$). Some show long term modulations, e.g. ESHC\,1, 2, 3,
4. The maximum difference in brightness over the whole time-span of observations
ranges between 0.2 and 0.45 magnitudes in $V_E$ (see Table\,\ref{lcchar}). Some
ESHCs show features in their light curve which maybe best termed by outbursts: an
increase in brightness is followed by a slow relaxation to the original flux
level. In particular, the observations of ESHC\,4 indicate (possibly) four outbursts. 
Two outbursts (at days 2485 and 2935) seem to be well covered by the observations. They 
last for 90 and 45 days respectively. In both cases the decrease towards
the initial flux level takes twice as long as the rising part. During the rise,
the colour of the object becomes increasingly redder (see Sect.\,\ref{discussion} for 
a discussion of the possible causes for the observed brightness and
colour behaviour). The same holds for a large outburst at
day 2503 of ESHC\,6. The brightness increase of ESHC\,3 at day 2921 seems to be
symmetric. The amplitude of the observed outbursts varies between $0.1^{m}$ and
$0.2^{m}$. The range in time-span of the observed outbursts in the different ESHCs
varies between 45 and 90 days. The outbursts with the largest brightness increases
seem to persist the longest.

In contrast, star ESHC\,5 seems to preferentially decrease in brightness from
the average flux level. This star displays variations on a somewhat shorter
time scale than the other ESHCs.  The colour of the object hardly changed during
the observations, while its brightness varied with $\Delta V_{\rm E}=0.25^{m}$.
Although to a lesser extent, ESHC\,6 might display decreases in brightness. This
is particularly observable from day 2800 onward.

During its first season of observation star ESHC\,1 shows quasi-regular
brightness variations on a $\rm \sim\,30 day$ time scale, superposed on a longer
time scale increase.  In the second and third season it is mainly at brightness
minimum. Star ESHC\,2 exhibits more pronounced long term modulation (see also
Beaulieu et al. 2001).

Examining the CMDs, one notices that for ESHC\,1, 2, 3, 4 and 6 the variations
are anti-correlated with the brightness of the star, i.e. it is
`redder-when-brighter'.  ESHC\,5 is an exception.  Quantitatively, the slope of
$d(V_{\rm E}-R_{\rm E})/dV_{\rm E}$ in a CMD should then have a negative value. We have
measured these slopes using a linear least-squares fit  to the average
colour in a magnitude bin of 0.02m containing at least 4 measurements (filled
symbols in the CMDs).  The values of this slope for ESHC\,1 to 6 are tabulated
in Table\,\ref{lcchar}. Only the slope of ESHC\,5 is compatible with grey
variation.

\subsection{The variability of ESHC\,7}
\label{lceshc7}
\begin{figure}
\centering
\includegraphics[width=6cm,height=8.5cm,angle=90]{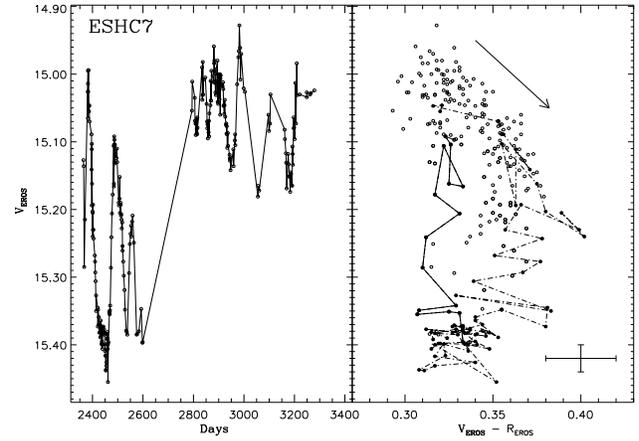}
 \caption{The light curve and CMD of ESHC\,7. The star first becomes redder when it fades
according to IS extinction by normal dust (indicated by the arrow for an $E(B-V)=0.10$). 
Then it reverses its trend and becomes `bluer-when-fainter'. The dashed and full lines indicate
consecutive measurements. For a description of the brightness-colour cycle see text.}
\label{eshc7cmd}
\end{figure}

ESHC\,7 (Fig.\,\ref{eshc7cmd}) has the largest amplitude of brightness
variations compared to the other 6 ESHCs.  The variations exhibit the same
trend, viz. a slow decrease followed by a sharp rise. Peculiarly, the deeper the
brightness minimum, the faster the subsequent increase is (for the deepest
minimum $\rm -0.018^{m}\,day^{-1}$ for the shallowest $\rm
-0.008^{m}\,day^{-1}$).  Part of the CMD of ESHC\,7  the faint part is
anti-correlated with its brightness (i.e. `bluer-when-fainter'), similar to
ESHC\,1, 2, 3, 4 and 6. However this is only the case when the star is in a
brightness minimum. The colour behaviour is different when the star is at
brightness maximum ($V_{\rm E}<15.2$).  Then the colour variations show the opposite
(i.e. `redder-when-fainter').

This behaviour is made more clear in the CMD. The dashed line connects the
consecutive measurements between day 2380 and 2480. During this phase the star
faded to its deepest observed brightness minimum.  Notice that during the
initial fading phase the star follows a `redder-when-fainter' branch. Remarkably
the slope of the measurements follow the expected slope arising from extinction
by normal IS dust (with $\rm R_{V}=3.3$). When the star has decreased its
brightness to $V_{\rm E}\sim 15.20$, a colour reversal occurs and the star follows a
`bluer-when-fainter' branch.

This reversal in colour behaviour (from redder to bluer) during the fading of
the star, is a well known and well studied behaviour of the UX\,Orionis subgroup
(UXOrs) of PMS stars.   It is understood in terms of obscuration by circumstellar
dust clouds, which causes the star to become redder when it fades initially. The
growing relative contribution of blue scattered radiation to the total observed
radiation due to the (unobscured) proto-planetary disk halts the reddening
colour trend, and makes it again bluer (Zaitseva
1987\nocite{1987Ap.....25..626Z}, Voshchinnikov et
al. 1988\nocite{1988Ap.....28..182V}, Eaton \& Herbst 1995, Grady et
al. 2000\nocite{2000prpl.conf..613G}, Natta et
al. 2000\nocite{2000prpl.conf..559N} and references therein).


However, contrary to what has been observed among UXOrs, during its
rise out of the first observed deep minimum, ESHC\,7 follows a third
and different branch. The star brightens at day 2463, but the colour does not
change. The corresponding measurements are connected by the full line in Fig.\,\ref{eshc7cmd}. 
It follows a track almost straight up, nearly
reaching its initial brightness. This cycle is partly repeated during
the second minimum at day 2535. The time the star spends in this minimum is shorter 
than during the first minimum. After this second brightness minimum, the observed flux increases 
at about constant colour, but in this case it does not reach its maximum
value. Consequently the star will join the colours corresponding to the
fading track, close to the position of the colour reversal. In Fig.\,\ref{eshc7cmd}
these are the measurements between $V_{\rm E}\sim15.35$ and  $V_{\rm E}\sim15.20$ and colour
$V_{\rm E}-R_{\rm E}\sim0.33$ and $V_{\rm E}-R_{\rm E}\sim0.35$ (see also Fig.\,\ref{fiteshc7}).

Summarizing, the brightness-colour cycle for ESHC\,7 consists of three
parts. Starting at brightness maximum: (1) a `redder-when-fainter' portion when
$V_{\rm E}\lta15.2$, (2) a `bluer-when-fainter' portion when $V_{\rm E}\gta15.2$ and (3)
a grey portion, to return to the initial brightness.  We will discuss this
behaviour in Sect.\,\ref{disceshc7}.

\begin{table*}
\centering
 \caption[]{The photometric measurements of the ESHCs.}
 \begin{tabular}{l|cccr|llll|lll}
 \hline
 \multicolumn{1}{c|}{} &
 \multicolumn{4}{c|}{EROS2} &
 \multicolumn{4}{c|}{ESO} &
 \multicolumn{3}{c}{2MASS} \\
 \multicolumn{1}{c|}{Name} &
 \multicolumn{1}{c}{$<V_{\rm E}>$}&
 \multicolumn{1}{c}{$<V_{\rm E}-R_{\rm E}>$}&
 \multicolumn{1}{c}{$\Delta V_{\rm E}$}&
 \multicolumn{1}{c}{$d(V_{\rm E}-R_{\rm E})/dV_{\rm E}$}&
 \multicolumn{1}{|c}{V}&
 \multicolumn{1}{c}{B-V}&
 \multicolumn{1}{c}{V-R}&
 \multicolumn{1}{c}{V-I}&
 \multicolumn{1}{|c}{J}&
 \multicolumn{1}{c}{J-H}&
 \multicolumn{1}{c}{H-K}\\
 \multicolumn{1}{c|}{(1)} &
 \multicolumn{1}{c}{(2)} &
 \multicolumn{1}{c}{(3)} &
 \multicolumn{1}{c}{(4)} &
 \multicolumn{1}{c}{(5)} &
 \multicolumn{1}{|c}{(6)} &
 \multicolumn{1}{c}{(7)} &
 \multicolumn{1}{c}{(8)} &
 \multicolumn{1}{c}{(9)} &
 \multicolumn{1}{|c}{(10)} &
 \multicolumn{1}{c}{(11)} &
 \multicolumn{1}{c}{(12)} \\
 \multicolumn{1}{c|}{} &
 \multicolumn{1}{c}{} &
 \multicolumn{1}{c}{} &
 \multicolumn{1}{c}{} &
 \multicolumn{1}{c}{} &
 \multicolumn{1}{|c}{} &
 \multicolumn{1}{c}{} &
 \multicolumn{1}{c}{} &
 \multicolumn{1}{c}{} &
 \multicolumn{1}{|c}{} &
 \multicolumn{1}{c}{} &
 \multicolumn{1}{c}{} \\
\hline
ESHC\,1 & 15.04 &  0.03 & 0.25 &    $  -0.24 \pm 0.02 $  & 15.06 & 0.00 & 0.00 & 0.04 & 15.05 & 0.09 & 0.13 \\
ESHC\,2 & 16.95 &  0.08 & 0.40 &    $  -0.33 \pm 0.03 $  & 17.00 &-0.06 &-0.01 & 0.03 &     - &    - & -     \\ 
ESHC\,3 & 16.61 & -0.01 & 0.20 &    $  -0.33 \pm 0.07 $  &  -    & -    & -    &  -   & -     & -    & -    \\
ESHC\,4 & 14.99 &  0.07 & 0.20 &    $  -0.16 \pm 0.04 $  &  -    & -    & -    & -    & 15.02 & 0.06 & 0.13 \\
ESHC\,5 & 13.98 &  0.19 & 0.25 &    $   0.04 \pm 0.07 $  &  -    & -    & -    &  -   & 13.48 & 0.12 & 0.09  \\ 
ESHC\,6 & 15.11 &  0.03 & 0.30 &    $  -0.08 \pm 0.03 $  &  -    & -    & -    &  -   & 15.02 & 0.07 & 0.34  \\
ESHC\,7 & 15.17 &  0.34 & 0.45 &        --               & 15.12 & 0.31 & 0.25 & 0.57 & 14.36 & 0.16 & 0.25  \\ 
\hline
 \end{tabular}
\label{lcchar}
\end{table*}

\section{H$\alpha$ emission and cross identification with SIMBAD}
\label{simbad}
Herbig Ae/Be stars are (partially) defined as having emission line spectra. Therefore we
systematically searched for cross identifications of the target stars with
archive catalogues using the SIMBAD database operated at Strasbourg, France.
Some ESHCs are reported as emission line stars in different SMC objective prism
surveys (Lindsay 1961\nocite{1961AJ.....66..169L}, Meyssonnier \& Azzopardi 1993
(MA)\nocite{1993A&AS..102..451M}).  MA classifies the observed H$\alpha$
profile and the observed continuum around the line. The H$\alpha$ emission of
ESHC\,5 is classified as ``moderate strength and peaked''.  The
H$\alpha$ line of ESHC\,7 is classified as ``moderate strength and
appreciably widened''.  For ESHC\,3 the detection of H$\alpha$ emission
is given as ``doubtful''.

\begin{figure}
\includegraphics[height=8cm,width=8cm]{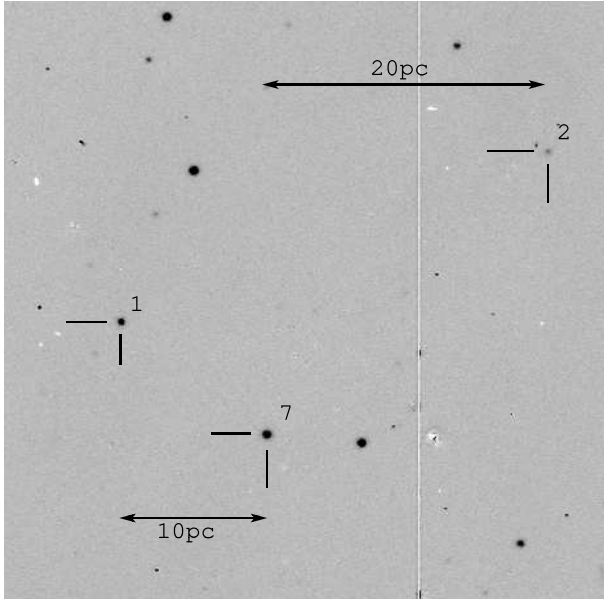}
 \caption{A subtracted $\rm H\alpha-H\alpha_{cont}$ image. The panel shows
  ESHC\,1, 2 and 7. Note that the stars are located within 35pc of each other.} 
\label{figha}
\end{figure}

The existence of H$\alpha$ emission is verified for 3 stars with the narrow band
imaging we obtained in 1998 (see Sect.\,\ref{danphot}).  We have built a
subtracted image of ($\rm H\alpha-H\alpha_{cont}$) using ISIS, the image
subtraction package with non constant kernel (Alard
2000\nocite{2000ilss.conf...50A}).  The resulting subtracted image is at 30\% of
the photon noise.  We show the resulting image in Fig.\,\ref{figha}. All
the objects in the image are H$\alpha$ emitters, the brightest of which have
been published before by Lindsay (1962) and Meyssonnier \& Azzopardi (1993).
The nature of these objects remains to be investigated, but they can be such objects as
PMS stars, PNe, \ion{H}{ii} regions, classical Be stars, symbiotic systems,
massive stars with winds.  We have indicated stars ESHC\,1, 2, 7. These stars
are clearly visible as H$\alpha$ emitters and are located within 35pc of each
other.

\section{The location of ESHCs in the SMC and their correlation with Far-IR emission}
\label{locelhc}
\begin{figure}
\includegraphics[height=8cm,width=8cm,angle=90]{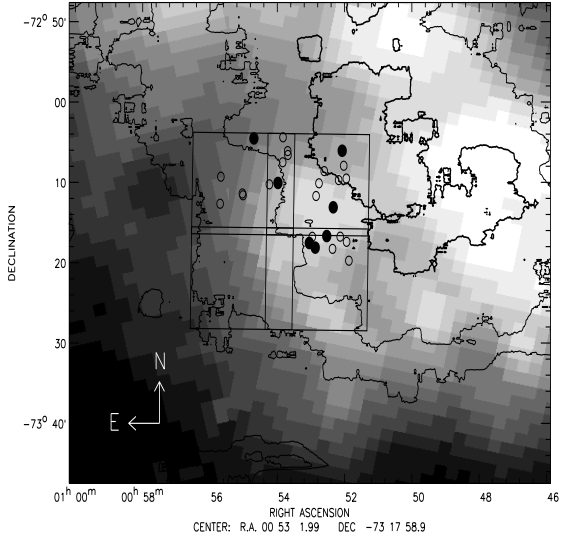}
 \caption{An IRAS image of the SMC at $\rm 60\mu m$. The FIR emission is 
        presented on a linear grey scale and 
        is in units of MJy/sr.  Indicated are the ESHCs (filled dotes),
        the blue variables (open circles). Overlaid are the
        fields corresponding to the EROS2 CCDs (large squares) and isophots of the SMC. Increasing thickness of
        the isophots means increasing stellar densities. Notice that
        5 ESHCs are associated with a slight enhancement of the
        $\rm 60\mu m$ emission.} 
\label{figIRAS}
\end{figure}

Regions with far IR (FIR) emission can be used as a tracer of massive
star formation. This was shown by Mead et al. (1990)\nocite{1990ApJ...354..492M}.
In Fig.\,\ref{figIRAS} we present a $\rm 60\mu m$ image taken with IRAS. It
shows the positions of the ESHCs (filled dots) in the bar of the SMC. The
contours represent stellar densities, which were determined from a Digital Sky
Survey image. Within the central contour (thickest
line) the highest stellar density is present. Note that the FIR emission does
not follow the distribution of stars. The maximum level of FIR emission 
(in MJy/sr) is visible at the far west in the image. Fig.\,\ref{figIRAS} shows
an enhancement of FIR emission in the central part.  There are 3 ESHCs (1, 2, 7)
grouped on the south edge of the enhancement and ESHC\,4 is on the east edge.
ESHC\,3 is located within the FIR emission region. The other 2 stars are located
outside the region. Its emission level is at 25\% of the maximum of
65\,MJy/sr. The lowest level of the IRAS image is at 5\% of the maximum.  The four
rectangles represent the EROS2 fields from Table\,1.

Fig.\,\ref{figIRAS} shows that both the HAeBe candidates and the blue variables
are generally (but not specifically) associated with the $\rm 60\mu m$ excess
emission, i.e. a region which is likely to have a young stellar population.  The
lack of any detection of blue variables in the lower right rectangle (which is
EROS2 field sm00103m) is noticeable, the more so, as the numbers of LPVs and
pulsating stars in this particular field are comparable to the other three EROS2
fields. Field sm00103m is the furthest from the central $\rm 60\mu m$
emission region compared to the other three examined EROS2 fields.

A correlation between FIR emission and location of HAeBe candidates in the LMC
was found by De Wit et al. (2002).  These stars were shown to be associated with
a $\rm 60\mu m$ excess emission region in the bar of the LMC.

\section{Deriving stellar parameters from SED fitting}
\label{sedfit}
In Sect.\,\ref{danphot} we described $BVRI$ and H$\alpha$H$\alpha_{cont}$
photometry for three fields in the SMC. These observations
were not specifically done on the fields in which the ESHCs are located.
However three ESHCs were within the field of view of these observations.  

In order to analyse the SED from visual to NIR, we performed a cross
identification with the 2MASS (2 Micron All Sky Survey) second incremental data
release (via http://irsa.ipac.caltech.edu). The limiting magnitudes of 2MASS
data for the detection of point sources with an $\rm S/N \gta 10$ are $J \le
15.8^{m}$, $H \le 15.1^{m}$ and $K_{s} \le 14.3^{m}$. We searched for sources
within 10'' of the target position. We checked the positions of the resulting
sources with the astrometrically calibrated optical images described in
Sect.\,\ref{danphot}. Available $\rm JHK_{s}$ measurements were found for the 5
brightest ESHCs. Thus, we have coverage of the SED
from the visual to the NIR for 3 ESHCs c.f. 1, 2 and 7. All photometric
measurements are listed in Table\,\ref{lcchar}.


We compared the $(B-V)$, $(V-I)$ and $(V-R)$ colours of the ESHCs with model
predictions to derive the effective temperature and/or the CS extinction. 
We chose not to use the $\rm JHK_{s}$ measurements in the fitting procedures. 
We used Kurucz models calculated for the appropriate metallicity for
the SMC, $\rm [Fe/H]=-1.0$. A minimal foreground extinction of 0.07
in $E(B-V)$ was applied, using the extinction law with $R_{v}=3.3$ from Cardelli,
Clayton \& Mathis (1989)\nocite{1989ApJ...345..245C}.  
In case of ESHC\,1 and 2 the stellar parameters have been derived from
spectroscopy by Beaulieu et al. (2001), viz. ESHC\,1: $\rm log(T_{\rm eff})=4.20$, $\rm
log(L/L_{\odot})=4.23$, ESHC\,2: $\rm log(T_{\rm eff})=4.34$, $\rm
log(L/L_{\odot})=3.84$. We used the constraints from their intrinsic colours to
derive the CS extinction by fitting for a minimum $\chi^{2}$.
For ESHC\,1 the atmosphere models corresponding to their effective temperature did not give 
a satisfactory fit to the data.
We obtained $\chi^{2} = 4.9$ for the best fitting model with an $E(B-V)_{cs}=0.06$.
For ESHC\,2 we obtained a $\chi^{2} = 2.4$ for an extinction $E(B-V)_{cs}=0.09$. 
In case of ESHC\,7, we do not have an independent determination of 
$T_{\rm eff}$ and $\rm log(L/L_{\odot}$). Therefore we fitted for both the temperature and
circumstellar extinction. The SED of ESHC\,7 fits a model atmosphere
of $\rm log(T_{\rm eff})=3.94^{+0.06}_{-0.06}$ with a $\chi^{2} = 1.7$. 
The corresponding CS extinction is $E(B-V)_{cs}=0.19\pm0.07$.  
The derived temperature of ESHC\,7 indicates an A-type star. 
We calculated its absolute visual magnitude, i.e. $M_{V}=-4.74^{+0.20}_{-0.20}$ with the 
SMC distance modulus of 18.94 (Laney \& Stobie 1994\nocite{1994MNRAS.266..441L}).

We adopt the bolometric correction of -0.184 
which follows from the best fitting Kurucz model. This results in 
a $\rm log(L/L_{\odot})=3.87^{+0.14}_{-0.18}$. The uncertainty in 
the derived luminosity of ESHC\,7 is estimated from 
the uncertainty in the effective temperature (bolometric correction) 
and extinction from the model fit, allowing all models which are within the 
internal error ($\sum_{i=1}^{N} 1/\sigma_{i}^{2}$) of the photometric measurements.

With respect to the EROS coverage of the light curves of the ESHCs, the
optical photometric measurements were obtained at day 2929 of Figs.\,4, 5, and
6, whereas 2MASS observed the ESHCs on 9sep 1998, corresponding to JD 2451034.5
or day 3142 in the same figures. Hence we see that for ESHC\,1, 2 and 7 the
optical and NIR photometric data for each object were taken during nearly equal
brightness states. Obviously, there is a dependence of the determined parameters
on the variability. For ESHC\,1 and 2, the derived extinction will be larger
when the stars are in a brighter state. For ESHC\,7, the photometric measurement
correspond to a high brightness state, in which the colour of the star is most
blue. The derived stellar parameters for ESHC\,7 may change when fitting
broad-band UBVI measurements taken at a different epoch, depending on the
physical mechanism causing the variability.

\begin{figure}
\centering
 \includegraphics[width=6cm,height=8cm,angle=90]{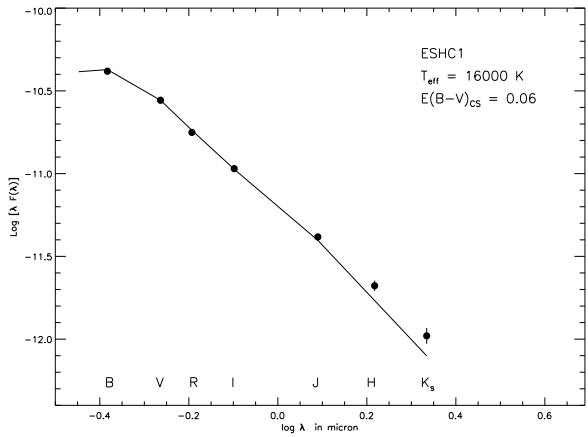}
 \caption{Spectral energy distribution for ESHC\,1. The NIR points are 2MASS
 data. The full lines are Kurucz atmosphere models for a  metallicity $\rm [Fe/H]=-1$.}
\label{eshc1sed}
\end{figure}

\begin{figure}
 \includegraphics[width=6cm,height=8cm,angle=90]{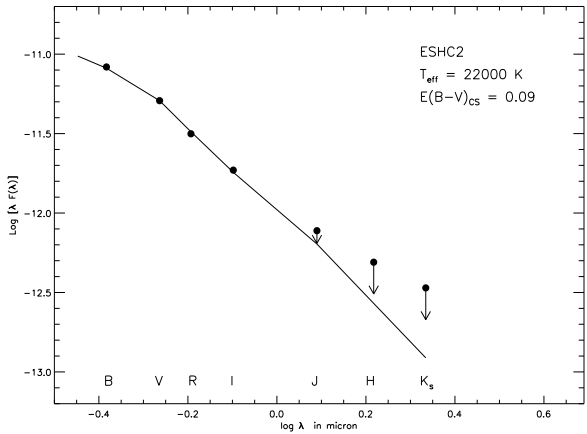}
 \caption{Spectral energy distribution for ESHC\,2. Upper limits for a S/N=10 from 2MASS are indicated.}
\label{eshc2sed}
\end{figure}

\begin{figure}
 \includegraphics[width=6cm,height=8cm,angle=90]{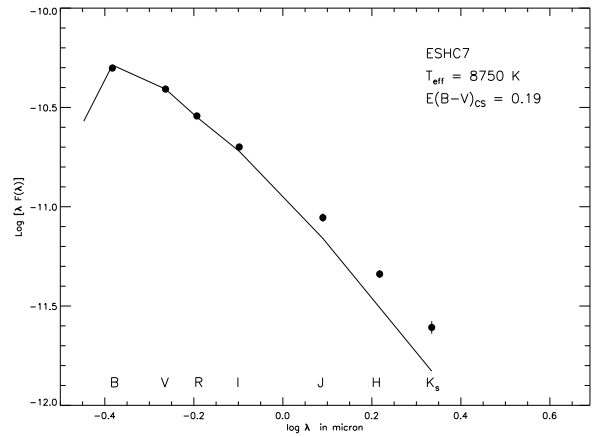}
 \caption{Spectral energy distribution for ESHC\,7. The NIR points are 2MASS data.}
\label{eshc7sed}
\end{figure}

The SEDs from B-filter to $\rm K_{s}$-filter of ESHC\,1, 2 and 7, with the
atmosphere models corresponding to the temperatures derived from spectroscopy
(ESHC\,1 and 2) and the SED fit (ESHC\,7) are presented in Figs.\,\ref{eshc1sed}
to \ref{eshc7sed}. The measurements of the stars are corrected for extinction
using the found values of $E(B-V)$, as described.  We conclude from these
figures that ESHC\,1 and 7 are brighter in $\rm JHK_{s}$ filters than expected
from stellar atmosphere models.


We will discuss the NIR excess in Sect\,\ref{nirprop}.

\section{Discussion of ESHC properties}
\label{discussion}
We have discussed  7 blue irregular variables in the SMC. The purpose
was to identify PMS stars at a lower than solar metallicity. 
Some of these stars (see Sect.\,\ref{locelhc}) are located near or on the edge of a region where an
increase in $60\mu $ emission has been detected by IRAS. FIR emission
has been shown to be a tracer of massive star formation (Mead et al. 1990)\nocite{}. We have
labeled them HAeBe candidates primarily on the basis of their brightness
variations, as an initial working definition. A number of criteria
exist to determine whether a star is a genuine member of the PMS
HAeBe group. Some criteria are degenerate with the characteristics
of classical Be stars, viz. H$\alpha$ emission, irregular
variability and near IR excess. In the next subsections we will 
discuss and compare these properties of ESHCs with the 
known properties of HAeBe stars and classical Be stars.

\subsection{Causes of ESHC brightness variability}
\label{causesofv}

Post-Main Sequence classical Be stars (B type stars of luminosity
class III to V, which show or have shown in the past the Balmer series
in emission) tend to be variable on several time scales and with a
range of amplitudes. A summary of light curves of Be stars over a 20 year period
have been published by Pavlovski et al. (1997)\nocite{1997A&AS..125...75P}.
Random brightness outbursts among early-type Be
stars have been measured by Hipparcos and discussed by Hubert \&
Floquet (1998)\nocite{1998A&A...335..565H}. These occur on time scales
between 20 to 500 days with $\Delta Hp\lta0.3^{m}$. The time scale of
the outbursts depends on the amplitude. On short-term time scales ($\rm
\lta days$) Be stars vary predominantly periodically with moderate
amplitudes, $\Delta V<0.1^{m}$ (Feinstein \& Marraco 1979\nocite{1979AJ.....84.1713F}). These
short-term time scales could either be caused by
non-radial pulsations or by rotation modulations (see Baade \& Balona 1994\nocite{1994IAUS..162..311B}).  
The large amplitude, long-term outbursts can be interpreted as variable
emission from the hot gaseous circumstellar environment (Waters et
al. 1987\nocite{}, Dachs et al. 1988\nocite{}). In this
case, the additional red flux from the disk will cause the brightness
to increase, and at the same time the observed SED to become flatter. 
The combined flux of star and disk will increase but will have 
a redder colour (e.g. Dachs 1982\nocite{1982IAUS...98...19D}). 

Pre-main sequence Herbig Ae/Be stars exhibit a large range of types in
photometric and spectroscopic variability. Generally, brightness
levels of HAeBe stars seem to be quasi-periodic, on time scales of
$\rm \sim year$ with $\Delta V\sim0.1^{m}$ (Herbst \& Shevchenko 1999\nocite{}). 
Superposed are random brightness decreases, which can have
large amplitude $\Delta V \gta1^{m}$.  The early type Herbig Be stars ($\rm
<B8$) however, tend to be low amplitude irregular variables with
$\Delta V<0.2^{m}$ (Finkenzeller \& Mundt 1984\nocite{1984A&AS...55..109F}, Van
den Ancker et al. 1998\nocite{1998A&A...330..145V}, 
Herbst \& Shevchenko 1999).
Obscuration of the central stars by a varying amount of circumstellar
dust is the most quoted mechanism to explain these variations (see
Wenzel 1969\nocite{1969IAUComp61}, Th\'{e} 1994\nocite{1994nesg.conf.....T}). Although unsteady mass accretion as
the dominant cause has also been proposed by Herbst \& Shevchenko (1999).

We will explore the brightness and colour variations exhibited by the 
ESHC stars in terms of the two physical processes quoted for 
Pre-MS and Post-MS Be star variability: (1) bound-free and free-free (bf-ff) gas emission, (2) 
variable dust obscuration. Both mechanisms are capable of causing 
a `redder-when-brighter' behaviour. In case (1)
this is caused be an increase in the total flux, while in case (2) it
is required that an extensive blue scattering source is present, which
will dominate when the total stellar flux decreases (and thus in fact `bluer-when-fainter').

In the next subsection we will first determine the sensitivity
of variable H$\alpha$ emission on the EROS2 broad filters.  This is
important as both HAeBe stars (e.g. Catala 1999\nocite{1999A&A...345..884C},
Pogodin et al. 2000\nocite{ 2000A&A...359..299P}) as well as classical Be stars (inherent to their
definition) are known to exhibit variations in the  equivalent width (EW) of H$\alpha$. 
 For Be stars an  H$\alpha$\, variability of 50A seems to be
maximum (see Dachs 1988, Banerjee et al. 2000). Whereas an HAeBe star like
\object{HD\,200775} can increase its EW(H$\alpha)$ from 55\AA\ to 105\AA\ (Pogodin et
al. 2000).

\subsubsection{The influence of H$\alpha$}
\begin{figure}
\centering
 \includegraphics[width=6cm,height=6cm,angle=90]{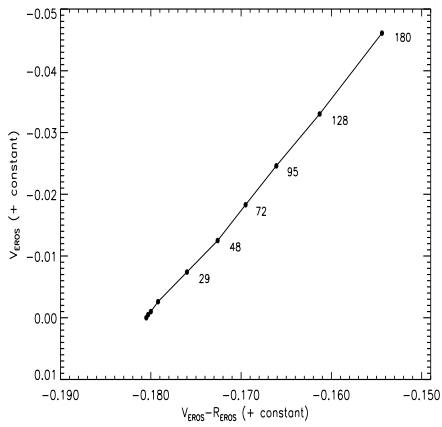}
 \caption{The correlation between colour as
 function of $V_{\rm E}$ magnitude. The corresponding $\rm H\alpha$ equivalent
 widths (in \AA) are indicated.} 
\label{figVecol}
\end{figure}

A variable EW of the H$\alpha$ line influences $V_{\rm E}$ and $R_{\rm E}$ passbands
differently and hence the colour will change as a function of the H$\alpha$ EW.
We will determine the qualitative contribution of H$\alpha$ to the EROS2
photometry.  An increasing amount of flux was added to the H$\alpha$ line of a
synthetic B type spectrum of $T_{\rm eff} = 23\,000{\rm K}$. At each increase we
folded the $V_{\rm E}$ and $R_{\rm E}$ response curves with the spectrum in order to
calculate the subsequent altered magnitudes and colour.  In Fig.\,\ref{figVecol}
we illustrate the dependence of $V_{\rm E}$ on H$\alpha$ EW and the subsequent
change in $V_{\rm E}-R_{\rm E}$.


We note that the influence of variable
H$\alpha$ emission on the magnitudes is small.  Its influence on
the total flux in the $V_{\rm E}$ filter is 4.4\% at an EW level of
180\AA, while it is 7.0\% on $R_{\rm E}$.  This leads to
a minor increase of $\sim 0.05^{m}$ and $\sim 0.08^{m}$ in
magnitude, respectively.  The observed maximum amplitude of variation
of the ESHCs, listed in Table\,\ref{lcchar}, is much larger.
However their small amplitude variations are of the same value as predicted
for strong H$\alpha$ variation. 

We can calculate the necessary EW variations in the case of ESHC\,1
and 2, if all the observed variations were caused by H$\alpha$ variations.
Beaulieu et al. (2001) measured an H$\alpha$ EW of $\rm 2.6\AA$ and $\rm 16.5\AA$,
respectively.  ESHC\,1 has an amplitude of variation of $0.33^{m}$ in
$V_{\rm E}$.  Then the EW should increase to $\rm 1300\,\AA$ to cause $\Delta
V_{\rm E} = 0.33^{m}$, hence an increase of almost 500 times. For ESHC\,2
the required H$\alpha$ variation is less, a factor 100. 

Because the influence on $R_{\rm E}$ is larger than on $V_{\rm E}$, a change
in the EW of the H$\alpha$ line will cause the star to become bluer
when it fades.  The subsequent slope of $d(V_{\rm E}-R_{\rm E})/dR_{\rm E}$ in
Fig.~\ref{figVecol} equals $-0.58$. None of the measured slopes
of the ESHCs have this value. Therefore we conclude that the photometric variations
are not solely due to a variation of the H$\alpha$ equivalent width.

\subsubsection{The influence of variable bound-free and free-free emission}
\label{infbbff}
In this section we present a comparison of the observed brightness and
colour variations of the program stars with an outflowing disk
surrounding a classical Be star.  The method was presented by Lamers \& Waters 
(1984)\nocite{1984A&A...136...37L} for spherically symmetric stellar winds and
adapted by Waters (1986)\nocite{1986A&A...162..121W} for a non-spherically
symmetric configuration to explain the observed infrared
(IR) excess emission of Be stars in terms of mass loss. It assumes
bf-ff radiation from an isothermal, equatorial, gaseous
outflowing disk, which is viewed at most pole-on (henceforth to be called
ff-disk). In a subsequent paper Waters et
al. (1987)\nocite{1987A&A...185..206W} showed that the observed
`redder-when-brighter' behaviour in a (B-V)-V diagram of the Be star 
\object{$\rm \kappa$\,CMa} (Dachs et al. 1986\nocite{1986A&AS...63...87D}) can be
explained by an increase of the ff-bf radiation of
the equatorial ff-disk.  When the number density of charged particles
(parameterized by the emission measure $\rm EM = \int n_{\rm E}^{2} dV$) in
the ff-disk increases, the total flux radiated by the ff-disk will increase
more in the lower energy part of the spectrum than in the higher
energy part.

Lamers \& Waters (1984)\nocite{1984A&A...136...37L} derived an analytical
expression for the monochromatic excess flux ratio ($\Delta
F_{\nu}^{tot}/F^{*}_{\nu}$) as a function of the temperature of the outflowing material, the
stellar radius, and the EM. We will use the expression as it
was presented in Eq.\,15 in Waters (1986).  This equation is valid in the
optically thin case. It can be applied if the stellar parameters are known,
viz. for ESHC\,1, 2 and 7. We can calculate the increase in flux in the $V_{\rm E}$ and $R_{\rm E}$ filters
due to an increase of bf-ff emission and compare this to the observed 
photometric measurements of ESHC\,1 and 2.  For the calculation of 
the {\it monochromatic} excess
flux we first determined the effective wavelengths of the EROS2
filters as a function of spectral type.  We used blackbody spectra
with the effective temperatures of ESHC\,1 an 2.  For a B4 star (ESHC\,1)
$\lambda_{\rm eff}$ is $\rm 5880\AA$ and $\rm 7380\AA$, for $V_{\rm E}$ and $R_{\rm E}$
respectively.  The difference in $\lambda_{\rm eff}$ for a B2 star
(ESHC\,2) is marginal.

\begin{figure}
\centering
\includegraphics[height=8cm,width=6cm,angle=90]{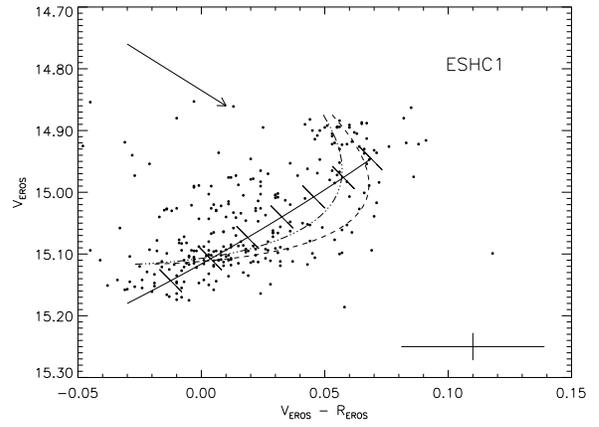}
 \caption{Colour-Magnitude diagram of ESHC\,1. The solid line represents the
optically thin ff-disk model (Sect.\,8.1.2). The tick marks on the model fit
indicates the increase in EM (to upper right corner) in steps of $\rm
0.2\,x\,10^{60}\,cm^{-3}$ to a maximum of $\rm 1.4\,x\,10^{61}\,cm^{-3}$. The
dashed lines represent obscuration model fits with an CS $R_{V}$ of 3.3 (right
one), the dot-dashed line has a CS $R_{V}$ of 5.0 (left one). Average error bar
is indicated in the lower right corner and a reddening vector (corresponding to
value of $E(B-V)=0.1$) in the upper left corner.} 
\label{figBedisc1}
\end{figure}

\begin{figure}
\centering
\includegraphics[height=8cm,width=6cm,angle=90]{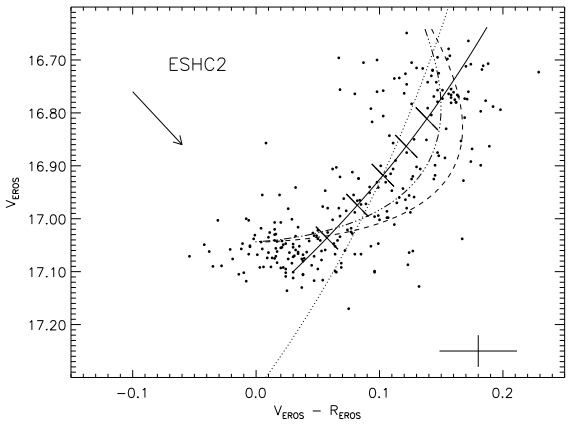}
 \caption{Same as Fig.\,\ref{figBedisc1}, but for ESHC\,2.  The tick marks on the
ff-disk model fit (solid line) indicates the increase in EM in steps of $\rm
1\,x\,10^{60}\,cm^{-3}$ to a maximum of $\rm 5\,x\,10^{60}\,cm^{-3}$ This model
fits if a CS extinction of $E(B-V)=0.12$ is applied. The extinction vector is
indicated by the arrow and corresponds to a value of $E(B-V)=0.1$. The dotted
line is an ff-disk model assuming that the total colour excess with respect to
the intrinsic colour is only due to ff-disk emission. The dashed lines represent
obscuration models for $R_{V}$ of 5.0 (left one) and 3.0.  The colour of the
applied scattering nebula is 0.04. An IS extinction was applied of
$E(B-V)_{IS}=0.05$. A reddening vector corresponding to $E(B-V)=0.1$ is
given. The figure is truncated on the $V_{\rm E}$-axis. } 
\label{figBedisc2}
\end{figure}

In Figs.\,\ref{figBedisc1} and \ref{figBedisc2} we show the CMDs of ESHC\,1 and
2. The solid lines  with tick marks represents the optically thin ff-disk model. Following
Waters (1987), we assumed that the temperature of the ff-disk is 0.8 times the effective
temperature of the star. We assumed that at brightness minimum the EM of the ff-disk is
negligible and we set its value to 0. The figures show that the observed relationship
for ESHC\,1 between visual brightness and colour is compatible with an outflowing
isothermal ff-disk, which increases its EM to $\rm 1.6 x 10^{61}\,cm^{-3}$, although
the scatter is large.  The maximum EM for ESHC\,2 is less, 
$\rm 7 x 10^{60}\,cm^{-3}$.

The model translates into an excess emission due to the contribution of the ff-disk. 
Therefore, the magnitude and colour of the B star itself is assumed to be
observed when the stars are at their faintest. These colours should correspond to
the intrinsic colours inferred from the spectral types derived by Beaulieu et al. (2001).  
From Figs.\,\ref{figBedisc1} and \ref{figBedisc2} we read that ESHC\,1 
at brightness minimum has $V_{\rm E}\sim15.15^{m}$ and $V_{\rm E} - R_{\rm E} \sim -0.03$, while 
ESHC\,2 has $V_{\rm E}\sim17.1^{m}$ and $V_{\rm E} - R_{\rm E} \sim 0.00$.

We converted the spectral types to the $V_{\rm E}-R_{\rm E}$ colour.
We adopted the $T_{\rm eff}$-$(V-I)_{0}$ conversion for 
SMC metallicity from Kurucz models. We applied the transformation equations 
from Sect.\,\ref{oglecal} and corrected for the IS extinction 
towards the SMC. Then we derive a $V_{\rm E}-R_{\rm E}=-0.04$ for ESHC\,1 and 
$V_{\rm E}-R_{\rm E}=-0.10$ for ESHC\,2.

There is a very good agreement between the derived spectroscopic colour 
of ESHC\,1 and its observed colour at its faintest state. The assumption
of a ff-disk causing the emission excess holds for ESHC\,1.

The derived spectroscopic colour and the observed colour in the faint state do not
agree for ESHC\,2. The spectroscopic colour is bluer. The discrepancy can be resolved in two ways.
Either (1) at its faintest phase the EM is non-zero, i.e. substantial emission from
the ff-disk is still present, causing the colour to be redder than the colour derived
from spectroscopy or (2) there is an additional CS extinction.  

Case (1) implies that the underlying star is intrinsically much fainter than the
minimal observed brightness of the EROS2 light curve. We calculated a ff-disk model
using the spectroscopic colour of -0.10. It is represented by the
dashed-dotted line in Fig.\,\ref{figBedisc2}. In its faintest observed state, a ff-disk
is present. An EM of zero (no ff-disk) is found at a $V_{\rm E}$ magnitude of
17.6. This model does not agree with the observed photometric measurements.
The full line in Fig.\,\ref{figBedisc2} represent a star+ff-disk system which is
subject to a CS extinction of $E(B-V)_{CS}=0.12$. In this model it is assumed
that no ff-disk is present in the faintest observed state. This model
is compatible with the EROS2 observations. The resulting dereddened
magnitude of the underlying B star in this case is $V_{\rm E}=16.8^{m}$. 

We conclude that in the case of ESHC\,1 the `bluer-when-fainter' behaviour can
fairly well be explained by an increase in the EM of the ff-disk. During
some epochs, the ff-disk contribution is marginal and the intrinsic
colours of the star is observed. These colours correspond to the
colours derived from the spectral type, corrected for interstellar
extinction. This is compatible with the notion of the 'Be-phenomenon',
i.e. stars temporarily go through a Be phase. However the required
maximum EM is high. Galactic stars of the same spectral type have a
typical EM which is an order of magnitude less (Waters et~al. 1987).
Their detailed calculations give a maximum EM of
$7.9 \rm{x} 10^{59}\,cm^{-3}$ for a B4e type star, whereas a value of about 
$1.2 \rm{x} 10^{61}\,cm^{-3}$ is required to explain the excess emission observed in ESHC\,1.

The case of ESHC\,2 is more complicated. A CS extinction of
$E(B-V)_{CS}\sim0.12$ is needed in order to let the ff-disk model fit. The same
extinction value was found in the best fit of the SED (Sect.\,\ref{sedfit}). The
extinction should originate in a region outside the ff-disk.
Furthermore, at the blue end of the CMD, the $V_{\rm E}$ magnitude of
ESHC\,2 seems to be constant while the colour gets bluer. This
cannot be explained by the ff-disk model. Comparing the derived EM for ESHC\,2
with Galactic Be stars, we find that the maximum EM is higher than Be stars of
similar spectral type. Waters et al. (1987) derive a maximum EM (model inferred, 
based on IRAS measurements) of $2.5 \rm{x} 10^{60}\,cm^{-3}$ for a B2e type
star, whereas a value of about $6 \rm{x} 10^{60}\,cm^{-3}$ is required to explain the excess emission observed in  
ESHC\,2. The derived large EM could just be a selection bias due 
to our variability criterion. We selected Be stars displaying the largest 
variability amplitude.

The exhibited brightness colour relation of ESHC\,7 cannot be explained
by a ff-disk model only. The increase in flux originating in the ff-disk will always
make the star `redder-when-brighter', whereas we observe `redder-when-fainter'
and peculiar `grey' phases. Therefore we conclude that it is unlikely
that the variability of ESHC\,7 is caused by variable ff-disk emission.
We did not specifically fit this star with the Be ff-disk model.

 For the objects for which we did not fit with the ff-disk model
(i.e. ESHC\,3, 4, 5 and 6), the analysis of the light curves presented in 
Sect.\,\ref{descoflc}, shows that 
three of them (ESHC\,3, 4, 6) exhibit bursts of emission. This phenomenon is 
preferentially observed among the early type classical Be stars.  
During the outbursts the colour of the Be star gets redder with a
colour gradient of $\Delta (B-V)/\Delta V \sim -0.1$ to $-0.3$ (Dachs 1982\nocite{}).  
The measured $d(V_{\rm E}-R_{\rm E})/dV_{\rm E}$ is tabulated in Table\,\ref{lcchar}.
ESHC\,3 and 4 have a gradient which is similar to the gradient of classical Be stars.
The gradient of ESHC\,6 is smaller, but still compatible with a Be ff-disk.
The brightest star in the sample (ESHC\,5) displays a different light curve than 
stars 3, 4 and 6, moreover its colour gradient is 
compatible with grey extinction. Therefore the photometric properties of this
star will not be accurately described by a ff-disk. 


\subsubsection{The influence of variable dust obscuration}
\label{infludust}
Obscuration by a variable optical depth of an inhomogeneous
circumstellar dust distribution has been successful in explaining the
observed photometric and polarimetric behaviour of UXOrs (recently reviewed 
by Natta et al. 2000\nocite{}).
The stars in question are mostly Herbig Ae or T\,Tauri stars with large $\Delta V
> 2^{m}$ amplitude variations. These occur randomly in time by
screening of the stellar flux by circumstellar dust clouds.  During
these deep brightness minima they show the `bluer-when-fainter' effect
(e.g. Gahm et al. 1993\nocite{1993A&A...279..477G}, Grinin et
al. 1994\nocite{1994A&A...292..165G}, Eaton \& Herbst 1995\nocite{1995AJ....110.2369E}).  
The blue flux is ascribed to scattering by small dust particles of the unobscured part 
of the dusty CS-disk/envelope of the HAeBe (Grinin 1988\nocite{1988SvAL...14...27G}, 
Natta \& Whitney 2000\nocite{2000A&A...364..633N}).

Lamers et al. (1999) and De Wit et al. (2002)\nocite{2002A&A...395..829D} analysed the observed
`bluer-when-fainter' behaviour of Large Magellanic Cloud HAeBe
candidates (ELHCs). In contrast to Galactic UXOrs the colour of about $50\%$ of these
stars {\it always} becomes bluer when the star dims. They argue that
this is a resolution effect.  An unresolved blue scattering nebula is present
around the ELHCs (which was one of the original criteria of Herbig
1960).  The nebula is much larger than the star. In this case when the central star fades, the 
relative contribution to the total flux of the scattering nebula 
will increase and the colour of the star+nebula system will become bluer.
This is strongly suggested by the derived negative extinction for  
some ELHCs, as determined by Lamers et al. (1999). They compared
the intrinsic colours determined from the spectral type with the
observed colours at maximum brightness. The observed colours were bluer than the
intrinsic colours. This produces a negative extinction correction,
indicating the presence of an additional source of blue flux.

It is important to note that there exists a difference 
between the scattering regions as interpreted for galactic
HAeBes and ELHCs. For HAeBes the scattering region is
in the direct vicinity of the star-disk system (i.e. a dusty disk
atmosphere). In the case of ELHCs the scattering region is assumed to be located
relatively far away. The blue flux is interpreted coming from an extended 
scattering circumstellar cocoon comparable to the one existing around
the Galactic HAeBe star V380\,Ori.

Following the arguments of Lamers et al. (1999) and De Wit et al. (2002), we can
infer blue scattering nebulae surrounding the ESHCs from the observed `bluer-when-fainter'
behaviour. These nebulae are not resolved due to the resolution limitations of our ground
based observations, where 1 arcsecond corresponds to 0.3\,pc in the SMC.



A full description of a star-nebula system using broad band photometry is
tentative. Using the analytical model described in De Wit et al. (2002)\nocite{}
we can calculate zeroth order estimates of the brightness and colour of the
scattering nebula. In the following the term extinction covers all processes
which contribute to both true absorption and scattering.  The model consists of
three components (1) the star, (2) the occulting dust cloud and (3) the
scattering nebula which is assumed to be spherically symmetric.  The dust cloud
is approximately of stellar size, and varies only in optical depth. The
scattering nebula is much larger than the obscuring cloud. During an observed
brightness decrease, only the star is obscured by the dust cloud.  There is no
contribution to the flux by the occulting cloud itself.  The scattering nebula
does contribute to the absorption of the stellar flux in the line of sight,
because the nebula is assumed to be circumstellar.  In this way, the total
extinction of the stellar flux is then determined by the interstellar
extinction, the nebular extinction and the extinction due to the occulting dust
cloud. Then, the difference with the observed flux is due to the constant
(scattered) flux contribution of the nebula.  The basic principle is that both
the amount of scattered nebular flux and the amount of extincted stellar flux by
the obscuring dust cloud will determine the observed amplitude of variation.

As input observables for our model we use (1) the intrinsic colours determined
from the spectral type, (2) the amplitude of variation, and (3) the magnitudes
at maximum brightness.  We assume that at minimum brightness the obscuring cloud
is optically thick, i.e. completely blocking the stellar radiation.  In this
way, we obtain the monochromatic {\it scattering} optical depth of the inferred
nebula, using the observed amplitude of variation. This is expressed in Eq. (5)
of De Wit et al. (2002).  We assume that at maximum brightness the stellar
radiation is attenuated only by nebular absorption and not by the obscuring dust
cloud.  In this way, we obtain the monochromatic {\it absorption} optical depth
of the inferred nebula, using the spectral type.  Now we can determine the
(constant) nebular flux in $V_{\rm E}$ and $R_{\rm E}$. Using the optical depth of the
obscuring dust cloud as a free parameter, we can calculate the resulting colour
variation as a function of brightness of the star+nebula system.

Dust surrounding young stars can be of an anomalous nature. This affects the
selective to total extinction. Therefore in our calculation we treated
interstellar (IS) grains and the circumstellar grain composition (both obscuring
dust cloud and the scattering nebula) separately.  For IS dust we used
$R_{V}=3.3$, while we fitted the CS $R_{V}$ as a free parameter. Moreover, we
tried to find the best solution by varying the amount of IS extinction.  This is
of importance because the applied amount of IS extinction affects the
determination of the nebular absorption optical depth, and hence the final
nebular flux.

As illustration we present the case of ESHC\,1 and 2 in
Figs.\,\ref{figBedisc1} and \ref{figBedisc2} to facilitate comparison with
the Be star ff-disk model.  In both figures we plotted models with
different $R_{V}$ for the CS dust, namely 5 and 3.3. The model fits are
indicated by the dashed lines. The optical
depth of the obscuration cloud ranges from 0.2-7.5 for ESHC\,1 and
from 0.3-4.9 for ESHC\,2.  The characteristics of the nebula and the
model parameters are given in Table\,\ref{tabnebula}

\begin{table*}
\centering
 \caption[]{Model parameters and nebula characteristics calculated with the obscuration model. Quoted 
values are for $R_{V}=3.3$ (see text).}
  \begin{tabular}[t]{ccccccccc}
   \hline
    Name    & $V_{\rm E}^{neb}$ & $(V_{\rm E}-R_{\rm E})^{neb}$ & $E(B-V)_{IS}$ & $E(B-V)_{CS}$ & $\tau_{scat}^{Ve}$ & $\tau_{abs}^{Ve}$ & $\tau_{scat}^{Re}$ & $\tau_{abs}^{Re}$ \\
   \hline
   ESHC\,1  & 15.23 & 0.031 & 0.04 & 0.2-7.5 & 1.45 & 0.40 & 1.03 & 0.24 \\
   ESHC\,2  & 17.19 & 0.042 & 0.05 & 0.3-4.9 & 1.03 & 0.79 & 0.84 & 0.47 \\
   ESHC\,7  & 15.45 & 0.32  & 0.02 & 0.2-6.5 & 1.03 & 0.57 & 1.06 & 0.34  \\
   \hline
    \end{tabular}
\label{tabnebula}
\end{table*}

The obscuration model is compatible with the observed colour
variations. Especially, in the case of ESHC\,1, the cluster of points at the
brightest phase corresponds to the predicted point at which the blue nebular
flux starts dominating the total flux.  However the ff-disk
model seems a better fit to the measurements of ESHC\,1. Calculating a formal
$\chi ^{2}$ value for the fits, we find that with a value of 18.2 for the ff-disk
and a 21.7 for the obscuration model with $\rm R_{v}=5.0$, the ff-disk fits
better.

The near constant brightness of ESHC\,2 at its faint phase while the colour
becomes bluer can be explained with this model as the nebular flux dominating at
this epoch (although the scatter of the measurements is large). Such a behaviour
can not be explained by the Be ff-disk model.  Quantitative comparison using
$\chi ^{2}$, shows a formally better fit of the obscuration model with $\rm
R_{v}=5.0$ then the ff-disk fits, 18.8 against 19.8. We emphasize that the
derived values for $\chi ^{2}$ in case of ESHC\,1 and 2 are statistically
insignificant, however we want to show the trends of the colour-magnitude
behaviour of the ESHCs in direct comparison to two possible models.

\subsubsection{The case of ESHC\,7} 
\label{disceshc7}
\begin{figure}
\includegraphics[height=8cm,width=6cm,angle=90]{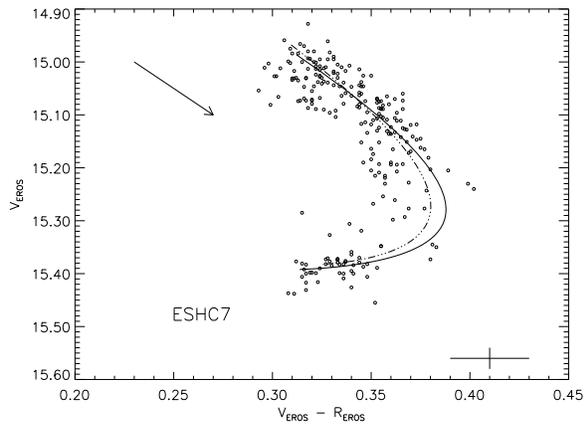}
 \caption[]{Dust Obscuration model fit for ESHC\,7. The colour of the nebula is 0.32. An IS extinction
 was applied of $E(B-V)_{IS}=0.02$. The thick line represents a model with an CS
 $R_{V}$ of 3.3. The dotted line has an $R_{V}$ of 5.0. For clarity purposes we
 omitted in the figure the measurements corresponding to the grey part of the 
 brightness behaviour of ESHC\,7. The reddening vector corresponds to an $E(B-V)=0.1$.}  
\label{fiteshc7}
\end{figure}

ESHC\,7 is most likely an Ae type star, which displays distinct colour
variability. As mentioned in Sect.\,\ref{lceshc7} this type of variations is
most commonly seen in PMS UXOri stars. However there are some differences
between the characteristics of UXOr variability and that displayed by
ESHC\,7. As far as literature goes, a `grey' phase as observed in ESHC\,7 has
never been observed among UXOrs. The minima of UXOrs are asymmetric, the star
recovering more slowly than it fades (Eaton \& Herbst 1995\nocite{}). ESHC\,7
shows the opposite trend in which the fading part lasts longer. Moreover the
time scale and amplitude are somewhat different.  UXOrs can experience a deep
($\Delta V >2$) brightness minimum and subsequent recovery to the initial
brightness within typically 50 days. In the case of ESHC\,7 it takes 100 days
to complete its deepest minimum and subsequent recovery to initial brightness of
$\Delta V_{\rm E}=0.4$.  At least the difference in amplitude can be understood with
the assumption of an unresolved and {\it bright} scattering nebula.

Dust is likely to play a role in the mechanism which causes the observed
variations.  This is suggested by the way the star changes its colour during
variations at bright phases. At bright phases the colour change complies with
what is expected from dust extinction.  Therefore we fitted the
brightness-colour relation with the dust obscuration model.  The result is
exemplified in Fig.\,\ref{fiteshc7}. The `grey' phase has been omitted from the
figure. The subsequent model parameters are listed in
Table\,\ref{tabnebula}. Notice that the scattering efficiency of the nebula is   
equally efficient in $R_{\rm E}$ as in $V_{\rm E}$. This is consequence of the fact that at
brightness minimum it is assumed that the colour is dominated by the scattering
nebula. This colour is similar to the observed colour at maximum brightness. Because
the colour of the nebula does not change during the obscuration of the central
star, the nebula would be observed bluer than the reddened star.
A similar conclusion can be drawn from the observed variation of
UXOri in the V-(V-I) plane (Grinin et al. 1994\nocite{1994A&A...292..165G}).
The star+nebula model gives a reasonable fit to the observed behaviour of
ESHC\,7, although a rather low IS extinction had to be applied of
$E(B-V)_{IS}=0.02$.


 We can estimate the Keplerian orbit of the cloud and compare it to
the dust destruction radius. A simple estimate of the dust destruction 
radius assuming dust in thermal equilibrium with the stellar radiation
field and a dust destruction temperature of 1500\,K 
(Rowan-Robinson 1980\nocite{1980ApJS...44..403R}), 
gives a distance of $\rm ~14\,AU$ for ESHC\,7. The brightness decrease of
the first main minimum took $\rm \sim45\,days$. Knowing the radius of the star,
and assuming (1) the size of the cloud is on the order of the size of the star,
(2) a mass for ESHC\,7 of $\rm 2.5M_{\odot}$, and (3) that $\rm \sim45\,days$ is
the time scale it takes for the dust cloud to cover the star, one can calculate
the cloud velocity and using Kepler's law, its distance to the star. One obtains
a distance of $\rm ~12\,AU$. Comparing this to the dust destruction radius it 
could be that the obscuring dust is too close to the star and therefore
destroyed. As the smaller dust particles are generally hotter than the larger ones,
they will be destroyed first. This could lead then to the observed
grey-when-brighter behaviour of ESHC\,7.

\subsection{Near Infra Red properties}
\label{nirprop}
The near-infrared (NIR) excess emission of HAeBe stars is one of the
characteristics which defines this group. 
Importantly, the NIR colours form a diagnostic to distinguish HAeBe
stars from post-main sequence classical Be stars (see e.g.
Finkenzeller \& Mundt 1984 \nocite{1984A&AS...55..109F}).  The bf and
ff emission present in Be ff-disks causes a specific NIR excess which is different 
than the thermal dust emission present in the CS environment of
Galactic HAeBe stars. The total observed NIR excess depends
on the kind of dust, its temperature and on its abundance.

\subsubsection{Metallicity effect on NIR dust emission}
\label{meteffect}
Dust abundance is likely to depend on the metal abundance.  This may be
evidenced by the fact that in the SMC the visible interstellar extinction per
unit H column density is about an order of magnitude lower than in the Milky Way
(Bouchet et al.  1985\nocite{1985A&A...149..330B}). Apart from the IS dust
abundance, the CS dust abundance of stars can be affected as well. Therefore the
observed NIR excess emission among stars like e.g. B[e], AGB, and HAeBes,
residing in a lower metallicity environment can be lower when compared to their
solar metallicity counterparts.

Here, we estimate the effect of dust abundances on the NIR emission of 
Galactic HAeBe stars. What would the NIR colours look like if the Galactic HAeBes 
were formed in an environment with the abundance of the SMC?
Both the CS dust emission as well as the CS  
extinction of the stellar photospheric radiation will be affected. 
We use the data for Galactic HAeBe stars given by Hillenbrand et al. (1992)\nocite{}.
Doubtful members of the group, i.e. \object{$\omega$\,Ori}, \object{MWC\,137}, \object{HD\,52721}, 
\object{HD\,53367}, \object{HD\,76534} were excluded from the analysis.
We assume that the derived visual extinction for the HAeBe is caused only
by the CS dust. The dust emission itself is not affected by CS extinction.
The metal abundance for the SMC is thought to be about one tenth of the solar value. 
To estimate the amount of CS dust of an SMC HAeBe star, we now scale simply
the dust amount of the Galactic HAeBe stars by a factor 10. 

First we derived the stellar emission in JHK band from (1) the observed visual
magnitude, (2) the extinction derived by Hillenbrand, and (3) the intrinsic
colours from Koornneef (1982)\nocite{1992ApJ...397..613H} appropriate for their
spectral types.  The derived stellar component was subtracted from the observed
total NIR emission, which results in the NIR flux due to dust in the JHK
bands. The dust flux then was reduced by a factor ten, given the expected NIR
dust emission at SMC metallicity.  Then we applied the equally scaled extinction
correction to the stellar flux.  On top of this we added the corrected NIR dust
flux. Finally we reddened the scaled JHK values for the interstellar extinction
towards the SMC. 

 The analysis gives first order estimates of the metallicity effect on NIR
emission of HAeBe stars. The underlying assumption in this estimate is that the
JHK emission of the CS disk of an HAeBe star is optically thin. Support that the
CS disk emission at these wavelengths cannot be very optically thick comes from
the interferometric measurements made by Millan-Gabet et
al. (2001)\nocite{2001ApJ...546..358M}. These authors favor the interpretation
that the CS dust is distributed in spherical envelopes with relatively low
optical depth. However Natta et al. (2001)\nocite{2001A&A...371..186N} have
constructed optically thick disk models with puffed up inner walls (where the
JHK emission is thought to reside) which match the data of Millan-Gabet et
al. (2001). Thus, it is currently unclear whether the optically thin scaling
proposed here is appropriate for HAeBe stars in a low metallicity environment.

Another effect that is likely to influence the amount of CS dust, but which we
did not take into account is that due to a lower dust column density there will
be an increase in dust destruction by UV photons. 

\begin{figure}
\includegraphics[height=8cm,width=6cm,angle=90]{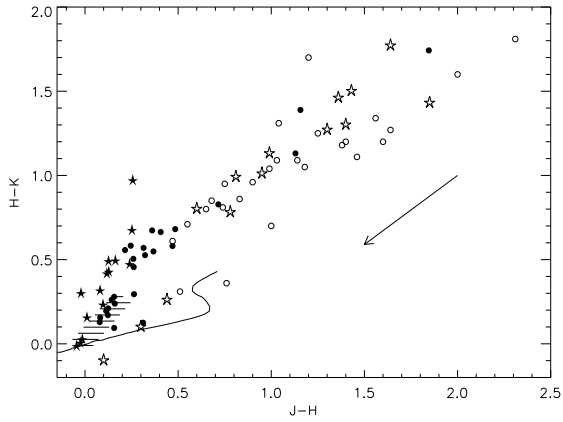}
 \caption[]{An SMC HAeBe colour-colour diagram. Each star is represented twice by
 (1) its observed Galactic colours and (2) its calculated colours for SMC metallicity. 
  The Galactic colours are presented by open symbols, the SMC colours by filled symbols.
  A distinction is made between Herbig Ae stars (circles) and Herbig Be stars (asterisks).
   This is a first order correction, under the assumption of optically thin NIR
  CS disk emission.
  Main Sequence colours for spectral type B1 to M4 are indicated by the 
  thick solid line (Koornneef 1983\nocite{1983A&A...128...84K}).
  The hatched area denotes the expected position of classical Be stars.
  The arrow signifies the correction in colour due to dust extinction with
 $R_{v}=3.3$, for an $A_{V}=5^{m}$.}
\label{jhhksmc}
\end{figure}

In Fig.\,\ref{jhhksmc} we present the expected influence of metallicity on the 
NIR colours of HAeBe stars. We display the resulting metallicity scaled $(J-H)$ and $(H-K)$ colour
of Galactic HAeBe by filled symbols. The original observed, Galactic NIR colours
are represented by open symbols. In the figure we make a
distinction between early type ($\rm <B8$, asterisks) and late type ($\rm \geq B8$,
circles) HAeBe stars for both the scaled NIR colours as well as the original
colours. The full line in Fig.\,\ref{jhhksmc} represents the Main-Sequence
adopted from Koornneef (1983). 

Metallicity affects the NIR emission profoundly. It will cause HAeBes to have a 
a small NIR excess. For three B3e stars, viz. RCW\,34, BD\,+65\,1637 and BD\,+41\,3731 
a negligible NIR excess results. However these 3 stars have an observed NIR excess
which is small.

The reduction in NIR excess emission of HAeBe stars due to the lower metal
abundance causes a degeneracy in  NIR properties between HAeBe stars and
classical Be stars. In Figs.\,\ref{jhhksmc} and \ref{jhhkeshc} we have exemplified the 
expected location of the Be stars by the hatched area. The Be star measurements are obtained 
from Dougherty et al. (1994)\nocite{1994A&A...290..609D}. Bound-free and
free-free emission is hardly affected by metallicity. The NIR colours of
some of the HAeBe stars are reduced to within the classical Be star regime.

We conclude that the NIR properties of low metallicity PMS stars could 
be different from the observed NIR properties of Galactic PMS stars, if these
properties are caused by optical thin thermal dust emission.

\subsubsection{Infrared properties of ESHCs}
\label{nireshc}
\begin{figure}
\includegraphics[height=8cm,width=6cm,angle=90]{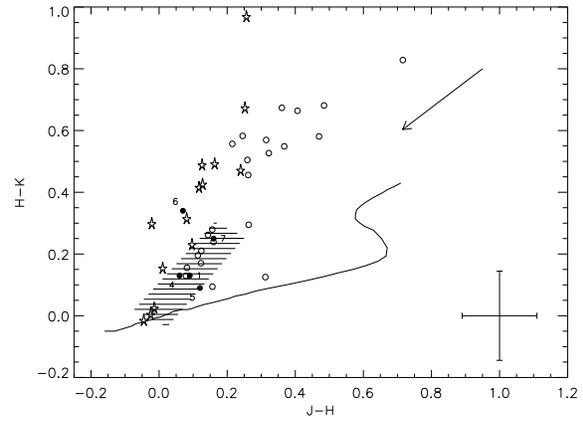}
 \caption{The Colour-Colour diagram for ESHCs (filled circles), with data from
 2MASS. The numbers of the ESHCs are indicated.  The open symbols are the
 Galactic HAeBe stars with scaled down dust content.  These points correspond to
 the filled symbols in Fig.\,\ref{jhhksmc}.  The arrow indicates the dependence
 of the colours on dust extinction, where its length represent a visual
 extinction of $A_{V}=2.5^{m}$. The figure shows that ESHC\,6 and 7 have
 substantial NIR excess (see Table\,4), comparable with the scaled Galactic
 HAeBe stars,while the NIR excess of ESHC\,1, 4 and 5 is minimal. The error bar indicated 
 is the one for ESHC\,6.}  
\label{jhhkeshc}
\end{figure}

We compare the observed NIR colours of the ESHCs with the expected colours of
Galactic HAeBe stars with scaled-down dust content in Fig.\,\ref{jhhkeshc}.
The expected metallicity-scaled NIR colours of Galactic HAeBe stars are displayed as open symbols.
We distinguish between the late type stars ($\rm \geq B8$) plotted as circles and the early type stars
plotted as asterisks, as in Fig.\,\ref{jhhksmc}. The ESHCs are symbolized by the filled circles.
The NIR emission of ESHC\,1, 4 and 5 is minimal with respect
to the scaled NIR emission of the Galactic HAeBes. 

The NIR emission of ESHC\,6 and 7 is relatively large. 
The observed NIR excess of ESHC\,6 is similar to the scaled Herbig B2e
star \object{HD\,259431}. 
This emission cannot be due to bound-free and free-free radiation.
The excess of ESHC\,7 is similar to scaled Herbig Ae stars like 
\object{$\rm LkH\alpha\,218$} and \object{HD\,150193}.

\section{Discussion of the nature of the ESHCs}
\label{discnat}
We have analysed the properties of 7 SMC stars. The aim was to identify HAeBe stars. 
We have probed different properties, which would corroborate the initial
working definition of {\bf E}ros {\bf S}MC {\bf H}AeBe {\bf C}andidates, viz. 
(1) $\rm H{\alpha}$ emission (2) spectral types (3) brightness and colour
variations (4) NIR excess emission (5) association with FIR emission.

\subsection{Summary of ESHC properties}
We will give brief a summary of the stars presented in this paper.
First the stars for which the evidence points to a particular evolutionary
stage. \newline
{\bf ESHC\,6} has an average $V_{\rm E}-R_{\rm E}=0.04$, which indicates a late to mid B
spectral type or earlier.  It has $\rm H\alpha$ emission. It exhibits both
outbursts and brightness decreases.  It has {a relatively large NIR emission 
but also with a large uncertainty in the $(H-K)$ colour. The emission could be 
caused by dust emission, but is not incompatible with bf-ff emission}. We
conclude that this star can be in a PMS phase. However it is not associated with
the $60\mu$ emission region.\newline
{\bf ESHC\,7} is a mid to early A type star with $\rm H\alpha$ emission. It shows
brightness variability very similar to PMS UXOri variability.  It has NIR excess
emission, which could be due to dust emission, but is not incompatible with
bf-ff emission. It is on the edge of the $60\mu$ emission region. This star is
likely to be in a PMS phase. 

For some ESHCs less data was available or contradictory evidence was found:\newline
{\bf ESHC\,1} is an early B type star with $\rm H\alpha$ emission. Both its NIR emission as well as the
colour gradient are compatible with bf-ff emission, which argues for a Post MS nature of ESHC\,1. 
However applying the Be ff-disk model to explain the brightness variations, the required EM 
is almost 20 times larger than observed among Galactic Be stars of the same
spectral type. 
The star is on the edge in the $60\mu$ emission region and is close to ESHC\,7. 
\newline
{\bf ESHC\,2} is an early B type star with $\rm H\alpha$ emission. Its NIR emission is unknown.
Its colour gradient has a value expected for bf-ff emission. The star is on the edge 
of the $60\mu$ emission region and is close to ESHC\,7.\newline
{\bf ESHC\,3} has an average colour of $V_{\rm E}-R_{\rm E}=-0.03$, which indicates a mid to early B spectral type
or earlier. It has $\rm H\alpha$ emission. Its brightness variability shows signs of outbursts. 
Its colour gradient indicated bf and ff emission. Its NIR emission is not known.\newline
{\bf ESHC\,4} has an average colour of $V_{\rm E}-R_{\rm E}=0.05$, which indicates a late to mid B spectral type or earlier. 
It is unknown if it has $\rm H\alpha$ emission. Its brightness variability shows strong 
signs of outbursts. It shows NIR emission compatible with bf-ff emission. Its colour gradient
has a value expected for bf-ff emission. This star is likely to be in a Post MS phase. \newline
{\bf ESHC\,5} has an average colour of $V_{\rm E}-R_{\rm E}=0.22$, which indicates a mid
to early A spectral type or earlier. It is an $\rm H\alpha$ emitter.  It is
$\sim 1^{m}$ more luminous than the other ESHCs.  Its colour variability is
nearly grey and it has a small NIR excess compatible with bf-ff emission.  Its
light curve shows both increases and decreases of brightness.  Judging from its
small NIR excess emission, the star might be a classical Be star. However in
that case, the star should have a CS extinction of $\rm E(B-V)\gta 0.2$, which
is large for a classical Be star (Waters et al. 1987).  Without proper
knowledge of the spectral type, we cannot make a preliminary determination of
the nature of ESHC\,5.

Note that our determination of the possible PMS nature of ESHC\,6 and 7 are
based on different criteria. In case of ESHC\,6, it is the NIR dust emission,
which forms the decisive argument. In case of ESHC\,7 the decisive argument is
its typical PMS brightness and colour behaviour.  One 
suspects ESHC\,6 to be a classical Be star based solely on its light curve.
This illustrates the need for an exact assessment of the difference between the
photometric behaviour caused by bf-ff emission and the proposed PMS mechanisms
like dust obscuration, wind variability, or accretion events. The more so
because we have shown in Sect.\,\ref{meteffect} that the NIR excess emission
among SMC HAeBe stars might be low, thus ceasing to be a defining criterion.

 On the other hand one can build a case for the fact that classical Be
star properties in the Magellanic Clouds are different from the ones in the
Galaxy. Studies have shown that the number ratio of Be/(B+Be) increases with
decreasing metallicity Maeder et al. (1999)\nocite{1999A&A...346..459M}. The
large derived EM for the ff-disk of ESHC\,1 and 2 in comparison with Galactic
Classical Be stars, may therefore be connected to differing Be properties with
metallicity. Indeed, a large sample of SMC Be star light curves from OGLE (among which
ESHC\,2) have been presented by Mennickent et al. (2002). The authors show
that many SMC Be light curves are different from the ones of Galactic Be objects.

\subsection{The location of the ESHCs in the HR diagram}
In this section we will determine the stellar parameters of the remaining ESHCs
(3, 4, 5, 6) for which 
neither spectrum nor SED is available. We use the EROS2 broadband
colours, the transformation equations deduced in Sect.\,\ref{oglecal},
and the the colours from model atmospheres by Kurucz with $\rm [Fe/H]=-1$.

We consider ESHC\,3, 4, and 5 as classical Be stars as discussed in the previous 
section. Therefore their intrinsic colours and magnitudes 
are best approximated when the star is in brightness minimum. We consider
ESHC\,6 as a HAeBe candidate. In principle the intrinsic parameters of an HAeBe star 
are obtained at brightness maximum. However the case of ESHC\,6 is peculiar. The star shows 
clear signs of outbursts. Moreover the star is in a low brightness state during 2 observing seasons.
It is unlikely that the star is obscured by dust clouds during this entire period. 
Therefore we choose the intrinsic magnitude and colour of this star observed at the epoch 
when it is brightest, outside the obvious outbursts. 

We apply an IS extinction of $E(B-V)=0.07$ and a distance modulus
of 18.94 as in Sect.\,\ref{sedfit}. We estimate that the cumulative 
uncertainty in the estimation of the intrinsic colour, due to 
transformation equations and measurements is 0.1 in (V-I).
The range in temperature is set by this uncertainty. The range in luminosity
is determined by the subsequent bolometric corrections, and an
adopted uncertainty in $V_{J}$ of 0.07, again due to the transformation to the
standard Johnson system and measurement uncertainties.
The derived values are tabulated in Table\,\ref{bepar}.
\begin{table*}
\centering
 \caption[]{Parameters of ESHC\,3 to 6, derived from EROS2 photometry, with an applied $E(B-V)_{IS}=0.07$.}
  \begin{tabular}[t]{crrrrrrrr}
   \hline
ESHC & $V_{\rm E}$ & $V_{\rm E}-R_{\rm E}$ & $V_{J}$ & $V-I$ & $M_{V}$ & $(V-I)_{0}$ & $\rm log(T_{\rm eff})$ & $\rm log(L/L_{\odot})$ \\
\hline
   3 & 16.69  & -0.04        & 16.68 & -0.04 & -2.49   &  -0.16   &$4.17^{+0.07}_{-0.09}$&$3.44^{+0.20}_{-0.21}$\\  
   4 & 15.09  &  0.05        & 15.07 &  0.07 & -4.10   &  -0.05   &$4.03^{+0.05}_{-0.08}$&$3.78^{+0.11}_{-0.19}$ \\
   5 & 14.14  &  0.22        & 14.20 &  0.32 & -4.97   &   0.20   &$3.90^{+0.02}_{-0.02}$&$3.93^{+0.03}_{-0.03}$ \\
   6 & 15.11  &  0.02        & 15.08 &  0.03 & -4.08   &  -0.09   &$4.08^{+0.07}_{-0.09}$&$3.87^{+0.14}_{-0.19}$ \\
   \hline
  \end{tabular}
\label{bepar}
\end{table*}
 
\begin{figure}
\includegraphics[height=12cm,width=8.5cm]{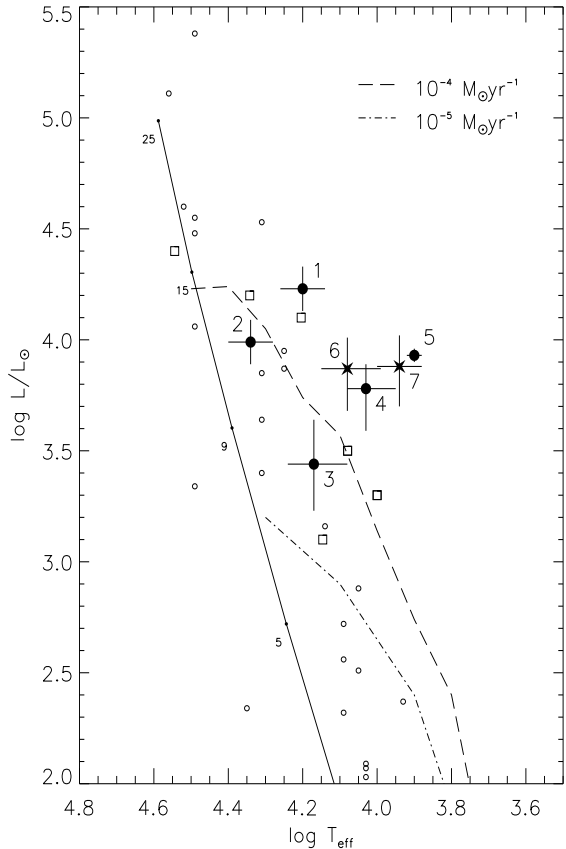}
 \caption[]{A HR diagram of the stars reported in this paper (filled
 symbols). Their numbers are indicated.
        The open squares are the LMC HAeBe candidates.
        Displayed are Galactic PMS stars (small circles) from Testi et al. 1998. \nocite{1998A&AS..133...81T}        
        The Main-Sequence with corresponding masses 
        is indicated by a full line. The dashed-dotted  and the dashed 
        lines indicate
        two predicted birthlines from Palla \& Stahler (1993, 1994)
        for a mean mass accretion rate of $\rm 10^{-5}M_{\odot}yr^{-1}$
        and $\rm 10^{-4}M_{\odot}yr^{-1}$ respectively.}        
\label{figHRD}
\end{figure}

In Fig.\,\ref{figHRD} we have plotted the stars described in this paper in an HR
diagram. The large filled symbols are the ESHCs presented in this paper.  The
two genuine HAeBe candidates, ESHC\,6 and 7 are presented by a filled asterisk.
ESHC\,1 and 2 are the two most luminous ESHCs. Their stellar parameters have
been determined from spectroscopy.  We added the LMC HAeBe candidates (ELHCs)
from Lamers et al. 1999, i.e.  the open squares. Their spectral types have also
been determined from spectra. The small circles are Galactic HAeBe stars,
adopted from Testi et al. (1998) \nocite{1998A&AS..133...81T}.  Note
however, that a PMS nature for the LMC HAeBe candidates has recently been
challenged by Keller et al. (2002).

We can compare the position of the clear HAeBe candidates ESHC\,6 and 7 with the
Galactic HAeBe population and the LMC HAeBe candidates. The two stars are
located in a region of the HR diagram where no Galactic HAeBe stars are
found. These SMC stars are 10 times more luminous than the Galactic HAeBe stars
of of the same spectral type.  Note that two of the Large Magellanic Cloud HAeBe
candidates (open squares) of the latest type are also more luminous than the
Galactic HAeBe population. However the figure suggests a difference between LMC
and SMC HAeBe candidates. Compared to a LMC star of the same spectral type
(ELHC\,6), the SMC star ESHC\,6 is nearly 2.5 times (0.4 dex) more
luminous. Star ESHC\,7 is almost 4 times (0.6 dex) more luminous than its LMC
counterpart ELHC5, which is of approximately the same spectral type. This might
be a first indication of a difference between the formation of intermediate mass
stars between the Galaxy, the LMC, and the SMC. However the numbers of genuine
candidates in the LMC and SMC are small and the stellar parameters of ESHC\,6
are based on colour information only.

The high luminosity of the ESHCs results in a discrepancy with the predicted
birthlines of Palla \& Stahler (1993,
1994)\nocite{1993ApJ...418..414P}\nocite{1994nesh.conf..391P} for the
Galaxy. This is also witnessed by the most massive Galactic HAeBe stars.  It is
important to note that the discrepant LMC and SMC stars are late B type or early
A type.  This may suggest a faster proto-stellar mass accretion rate at lower
metallicities. 

 On the other hand such a faster proto-stellar mass accretion rate may lead
to a more massive disk in the more evolved PMS phase and therefore to larger
IR excesses than the ones of HAeBe stars in the Galaxy. And as discussed in
Sect.\,\ref{nireshc}, this is not the case.  

Therefore, shedding more light on the nature of these objects would mean doing
deep IR photometry and narrow band H$\alpha$ photometry to search for the putative low mass
PMS stellar population near the ESHCs. In addition high resolution spectroscopy,
especially in the case of ESHC\,7, may lead to the detection of signatures of
the CS environment. For example, resolved H$\alpha$ profiles can be compared to
the expected profile from rotating CS disks/envelopes (double peaked), mass loss
(P\,Cygni), or a pole on viewed system (single peaked).

The profiles of emission lines can give information on the outflow 
velocity, which is typically between 200 and 500 km/s for Galactic HAeBe stars.
The comparison of the strength and velocity of the wind lines (H$\alpha$, \ion{Ca}{ii},
\ion{Na}{i}) of HAeBe stars in the Galaxy and SMC will help to understand the mass loss
mechanism and possibly reveal an accretion connection.


\section{Conclusions}
\label{conclusions}
(1) Two SMC HAeBe candidates (ESHC\,6 and 7) have been detected. They have $\rm
H\alpha$ emission, they are irregular variables, they have a NIR
excess.\newline 
(2) Five other SMC stars might be HAeBe candidates but their
properties are also compatible with Post Main-Sequence classical Be
stars.\newline 
(3) Metallicity can influence the expected CS NIR excess of HAeBe
stars in such a way that it should be used with caution as a discriminating
criterion between Pre Main Sequence Be stars and Post Main Sequence Be
stars.\newline 
(4) SMC HAeBe candidates might be more luminous than LMC or
Galactic HAeBe stars of the same spectral type. Whether this is due to a lower
dust abundance or a higher mass accretion rate is not clear. \newline


\begin{acknowledgements}
We would like to thank Dr Rens Waters and Dr Conny Aerts for useful
discussions. We thank the referee for his many valuable comments on the manuscript. 
The research was made possible through the use of the Simbad
database, operated at Strasbourg, France. This publication makes use of data
products from the Two Micron All Sky Survey, which is a joint project of the
University of Massachusetts and the Infrared Processing and Analysis
Center/California Institute of Technology, funded by the National Aeronautics
and Space Administration and the National Science Foundation.
\end{acknowledgements}

\bibliographystyle{aa}

\end{document}